\documentclass[onecollarge]{svjour2}       
\usepackage{graphicx}
\begin{document}

\title{Tidal friction in close-in satellites and exoplanets.}
\subtitle{The Darwin theory re-visited}
\author{Sylvio Ferraz-Mello \and Adri\'an Rodr\'{\i}guez \and Hauke Hussmann }
\institute{S. Ferraz-Mello and A.Rodr\'{\i}guez \at 
Instituto de Astronomia Geof\'{\i}sica e Ci\^encias Atmosf\'ericas\\
Universidade de S\~ao Paulo, Brasil \and
H.Hussmann \at
Institut f\"ur Planetenforschung, DLR, Berlin-Adlershof, Germany}
\titlerunning{Tidal friction in satellites and exoplanets}

\maketitle

\begin{abstract}
This report is a review of Darwin's classical theory of bodily tides in which we present the analytical expressions for the orbital and rotational evolution of the bodies and for the energy dissipation rates due to their tidal interaction. 
General formulas are given which do not depend on any assumption linking the tidal lags to the frequencies of the corresponding tidal waves (except that equal frequency harmonics are assumed to span equal lags).
Emphasis is given to the cases of companions having reached one of the two possible final states: (1)  the super-synchronous stationary rotation resulting from the vanishing of the average tidal torque; (2) the capture into a 1:1 spin-orbit resonance (true synchronization). 
In these cases, the energy dissipation is controlled by the tidal harmonic with period equal to the orbital period (instead of the semi-diurnal tide) and the singularity due to the vanishing of the geometric phase lag does not exist.
It is also shown that the true synchronization with non-zero eccentricity is only possible if an extra torque exists opposite to the tidal torque. The theory is developed assuming that this additional torque is produced by an equatorial permanent asymmetry in the companion. 
The results are model-dependent and the theory is developed only to the second degree in eccentricity and inclination (obliquity). 
It can easily be extended to higher orders, 
but formal accuracy will not be a real improvement as long as the physics of the processes leading to tidal lags is not better known.


\bigskip
\centerline{published in {\em Celestial Mechanics and Dynamical Astronomy} 101, 171-201 (2008). Incl. {\em Errata} in press.}
\bigskip

\end{abstract}

\def\beq{\begin{equation}}
\def\endeq{\end{equation}}
\def\begdi{\begin{displaymath}}
\def\enddi{\end{displaymath}}
\def\ep{\varepsilon}
\def\epp{\varepsilon^\prime}
\def\aprmaior{\;\buildrel\hbox{>}\over{\sim}\;}    
\def\defeq{\;\buildrel\hbox{\small def}\over{\,=}\;}    
\def\speq{\hspace{1mm} = \hspace{1mm}}    

\section{Introduction}

This report aims at presenting the main ideas of Darwin's classical theory of bodily tides (Darwin, 1879, 1880) and its consequences in the study of tidal friction effects on close-in satellites and exoplanets (hot Jupiters). In that sense, it is not original: the works of Goldreich (1963), Kaula (1964), Alexander (1973), Zahn (1977), Mignard (1979), Hut (1981), Eggleton et al. (1998), etc. already explored the consequences of Darwin's theory. They showed that Darwin's theory is sufficient to understand the main effects of tidal friction in the Solar System. Indeed, this report was initially written as an attempt to have a document presenting the fundamental equations of tidal friction in a simpler way, close to the approach followed by Jeffreys (1961). Kaula's very complete theory with consideration of higher order tides, and with many infinite series in eccentricity and inclination, must be handled with enormous care to avoid being lost in the successive summations and in the interpretation of some coefficients not unambiguously defined. An additional difficulty arises from the fact that many formulas appearing in the literature citing Kaula (1964) are not actually found in that paper. They may have been derived from those in the paper, but implicitly using additional assumptions (on lags, dissipation and rotation frequencies, for instance). These additional assumptions are never mentioned in these indirect citations. More simple and self-contained approaches to the problem are certainly useful. They are found in the literature but, most of them consider only parts of the subject and the existing results are fragmented in a large number of different papers. 
 
The main differences of this report with respect to Darwin's work concern the phase lags of the tidal waves. In Darwin's theory, each phase lag is assumed to be proportional to the corresponding wave frequency. Here, the main equations are obtained considering each lag as an independent quantity, only assuming that equal frequencies lead to equal lags and that the lag vanishes when the frequency goes to zero.
Results with assumptions on the lags are introduced afterwards, almost at the end of the paper, thus avoiding the need for reworking the whole theory when a different assumption is adopted. The equations given here for the orbit evolution do not depend on particular assumptions on the lags' dependence on frequencies.
Another important difference comes from the fact that Darwin (1880) considered simultaneously with the tidal effects those arising from the oblateness of the deformed body. In his theory, the orbital plane precesses due to the oblateness.

A major difference between this report and the traditional literature is the consideration, here, of the two different final states for the rotation of close-in companions. The first, called ``stationary rotation'', corresponds to the equilibrium situation reached by the close-in companion in which the average tidal torque acting on it vanishes and there is no other torque acting on the body. As it is well-known, if the orbital eccentricity of the companion is not zero, the resulting stationary state is a rotation slightly faster than the orbital motion (super-synchronous rotation). This is the situation adopted in the majority of studies of tidal evolution of planetary satellites and is expected to happen in the case of a fluid companion. The second, called ``synchronous rotation'', corresponds to the equilibrium situation reached when the companion has an important permanent, solid-body-like, equatorial bi-axiality ($J_{22}$ or $J_{31}$-component of its gravitational potential). 
In this case, tidal friction drives the rotation close to synchronous rotation allowing for the companion to be captured into a 1:1 spin-orbit resonance\footnote{We should add the possibility of capture into a different spin-orbit resonance long before approaching synchronization, as in the case of Mercury (see Lemaitre et al. 2006).} due to the torques arising from the equatorial bi-axiality. This is the situation adopted in the majority of studies of satellite rotation (Cassinian theories). It can be achieved by planetary satellites and close-in super-Earths.

In the study of orbital evolution, we preferred an approach different from that generally followed. Instead of using the direct Gauss or Lagrange equations to obtain the variation of the orbital elements, we preferred to obtain these variations from the application of conservation laws. 
The derivation is not as direct as with Gauss or Lagrange equations, but we consider it as important to keep in touch with the underlying physics. As a bonus, the equations giving the dissipation of energy in the interior of the deformed bodies are obtained as part of the process.

In addition, we mention that some studies consider only the force acting on one body due to the tidal deformation of the other and neglect the bulk reaction force acting on the deformed body (they consider only the torque due to that reaction). In most of such theories, this is not an error: They aim at studying the friction due to tides raised by a satellite on a planet, and the acceleration due to the reaction force has to be divided by the mass ratio planet-to-satellite. The effects are thus much smaller than the direct ones, but the results obtained with this assumption cannot be easily applied to different cases. 

Finally, it is also worthwhile mentioning that, following Darwin, we do not use Love theory (which did not exist when his works were published); it is noteworthy that the results usually obtained with Love theory do not essentially differ from those obtained just by adopting the first-order Jeans spheroid as equilibrium figure of the body under the action of an external attraction combined with Darwin's ideas on the lag and reduced height of tidal waves due to the visco-elastic nature of the bodies. One important  difference of the approach followed in this report with respect to many other works on Darwin's theory is the possibility of introducing different coefficients for harmonics of the tidal wave with different frequencies, instead of introducing the same ``Love number" for all harmonics arising from $U_2$ (see eqn. \ref{U2}). Anyway, the coefficients used in the theory are of the same nature as Love numbers and Love theory (Love, 1927) should be used to estimate their values (but such estimation is not done here).

This report is divided into many small sections.
Sections 2 to 7 include the necessary derivation of the forces between the two bodies due to the tide raised on one of them by the action of the other. In two of these sections (sections 3 and 4), the different tidal waves are described and the rules used to introduce tidal wave lags are presented. 
''Synchronous" and ``stationary" rotation of close-in companions are discussed in Sections 8 and 9. Sections 10 to 13 present the effects due to tides raised on close-in companions.  Section 14 presents the effects of tides raised on the central body. In section 15, the results are used to reproduce results from linear theories with a constant time lag. At the end, sections 16 to 18 discuss how the two tides must be added to obtain the variation of the orbital elements of the system. For sake of comparison to other works, section 16 gives the variation of the mean-motion and eccentricity under different hypotheses, with emphasis on linear theories with a constant time lag.

\begin{figure}[t]
\centerline{\hbox{
\includegraphics[height=4cm,clip=]{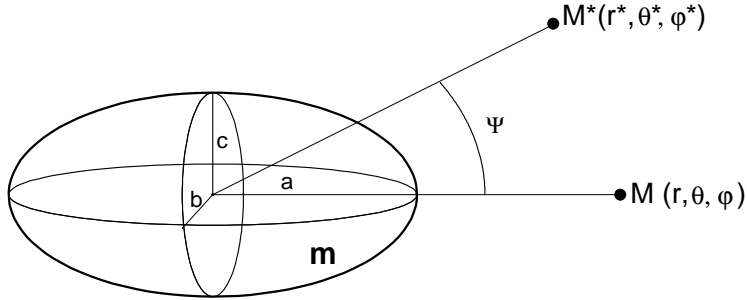}}}
\caption{Equilibrium figure of the deformed body $\tens{m}$ under the gravitational attraction of M}\label{esferoide}
\label{eplot1}       
\end{figure}

\section{The static equilibrium tide}

We consider two bodies orbiting one around another, separated by a distance $r$, and assume that one of the bodies is deformed under the tidal action of the other.
In the whole report, $\tens{m}$ is the deformed body and $\tens{M}$ is the outer mass (a mass point) responsible for its deformation\footnote {When establishing the main equations, we do not need to identify which is which in the system of two bodies. 
The relative size of the bodies may be considered only in the applications because the equations are valid even when the bodies have comparable masses.
To take into account the tides raised on both bodies, it is necessary to add the two effects following the equations given in the three last sections of the report.}.
We choose a reference frame with origin in the tidally deformed body and set $\mathbf{r}$ and $\mathbf{r}^{\ast}$ as the position vectors of, respectively, $\tens{M}$ and an arbitrary point $\tens{M}^\ast$ in space. 
The body $\tens{m}$ is initially considered as a homogeneous and perfect inviscid fluid that assumes the equilibrium shape dictated by its internal gravity and the tide generating potential due to the external mass $M$. 
If rotation is neglected, the equilibrium figure is a Jeans spheroid ($c=b$) whose axis of symmetry is pointed towards $\tens{M}$ (see figure \ref{esferoide}) and whose prolateness is 
\begin{equation}\label{prola}
\epsilon=\frac{a}{b}-1=\frac{15}{4}\Bigg{(}\frac{M}{m}\Bigg{)}\Bigg{(}\frac{R}{r}\Bigg{)}^3
\end{equation}
(Tisserand, 1891). $R$ is the mean radius of  $\tens{m}$. If the body rotates  and the external mass lies on the body's equatorial plane, the equilibrium figure changes into a Roche ellipsoid the smaller axis of which is directed along the rotation axis (see Chandrasekhar, 1969). The prolateness of the equator is the same as given above and it is, generally, almost invariant to the rotation velocity (see Appendix A). The forces arising from the polar oblateness may be, in a first approximation, superimposed on the tidal forces. 

\begin{figure}[t]
\centerline{\hbox{
\includegraphics[height=7cm,clip=]{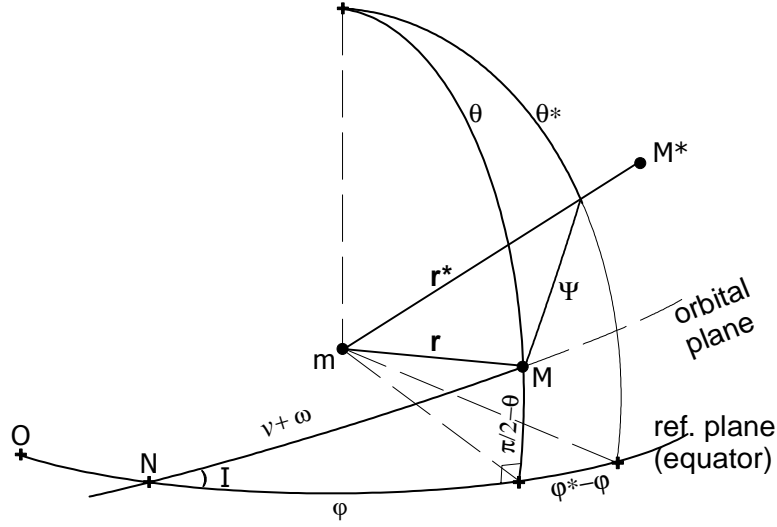}
}}\caption 
{Spherical coordinate system with origin at the center of mass of $\tens{m}$ and its equator as reference plane. The orbital plane is also shown.}\label{coord}
\end{figure}

The potential raised by a prolate spheroid of mass $m$ at an external point located in $\mathbf{r}^{\ast}$ is given by
\begin{equation}\label{Uini}
U=-\frac{Gm}{r^{\ast}}\Bigg{(}1+\frac{B-A}{2mr^{\ast2}}(3\cos^2\Psi-1)\Bigg{)}
\end{equation}
(see Beutler, 2005) where $A,B,C$ are the moments of inertia of $\tens{m}$ with respect to its principal axes.  ($A$ is the moment of inertia with respect to the symmetry axis of the spheroid; $C=B$.)
We note that, since the symmetry axis of the spheroid coincides with the direction of $\mathbf{r}$, $\Psi$ is the angle formed by the position vectors $\mathbf{r}$ and $\mathbf{r}^{\ast}$.
Assuming that $B$ is proportional to $a^2+c^2$ and $A$ is proportional to $b^2+c^2$, there follows $B-A\simeq\epsilon A$. Hence,

\begin{equation}\label{Uini2}
U=-\frac{Gm}{r^{\ast}}-\frac{15}{8}GA\Bigg{(}\frac{M}{m}\Bigg{)}\Bigg{(}\frac{R}{rr^{\ast}}\Bigg{)}^3(3\cos^2\Psi-1)
\end{equation}
or, introducing the parameter\footnote{This parameter is often called \emph{fluid Love number} because it is equal to the secular Love number of a rotating Maclaurin spheroid whose oblateness is $\epsilon$. For a homogeneous sphere, $k_f=1.5$ (since, in this case, the moment of inertia is $A=0.4\,mR^2$.)} $k_f=15A/4mR^2$,

\begin{equation}\label{U} 
U=-\frac{Gm}{r^{\ast}}-\frac{k_fGMR^5}{2r^3r^{\ast 3}}(3\cos^2\Psi-1);
\end{equation}
In order to express $\cos\Psi$ as a function of the components of the vectors $\mathbf{r},\mathbf{r}^{\ast}$, we choose a spherical coordinate system (figure \ref{coord}) so that $\mathbf{r}\equiv(r,\theta,\varphi)$ and $\mathbf{r}^{\ast}\equiv(r^{\ast},\theta^{\ast},\varphi^{\ast})$. The two angles considered for each point are their co-latitudes ($\theta, \theta^{\ast}$) and longitudes ($\varphi, \varphi^{\ast}$). We have

\begin{equation}\label{cosfi}
\cos\Psi=\cos\theta\cos\theta^{\ast}+\sin\theta\sin\theta^{\ast}\cos(\varphi-\varphi^{\ast}).
\end{equation}

Let us now introduce the relations between $r$, $\theta$, $\varphi$ and the orbital elements of $\tens{M}$. From the spherical triangle shown in the lower part of figure \ref{coord}, we have

\begin{equation}\label{sinteta}
\sin\theta=\cos(\omega+v)\cos\varphi+\sin(\omega+v)\sin\varphi\cos I          
\end{equation}

\begin{equation}\label{costeta}
\cos\theta=\sin(\omega+v)\sin I
\end{equation}
and
\begin{equation}\label{ecliptica}
\varphi\simeq v+\omega-\frac{1}{4}\sin(2v+2\omega)\sin^2 I+ O(I^4)
\end{equation}
where $\omega$ is the argument of the periapsis of $\tens{M}$ and $v$ is the true anomaly. $I$ is the obliquity, that is, the inclination of the orbital plane with respect to the reference plane (equator). From the equations of the Keplerian motion we have, in terms of the mean anomaly $\ell$, to second order in the eccentricity $e$:

\begin{equation}\label{fi}
v=\ell+2e\sin \ell+\frac{5}{4}e^2\sin 2\ell ,
\end{equation}

\begin{equation}\label{raster}
r=a\Big{[}1-e\cos\ell+\frac{1}{2}e^2(1-\cos 2\ell)\Big{]}.
\end{equation}
The next step is to substitute the above equations into (\ref{cosfi}) and (\ref{U}) and expand to $O(e^2)$ and to $O(I^2)$. For sake of simplicity we adopt the notations

\begin{eqnarray}
S&=&\sin I\\
P&=&\sin\theta^{\ast}\\
Q&=&\sin 2\theta^{\ast}
\end{eqnarray} 

At the orders considered, the static equilibrium tidal potential at the point $\mathbf{r}^{\ast}$ is

\begin{eqnarray}\label{u}
U_2&=&-\frac{3 k_f G M R^5}{4 a^3 r^{\ast 3}}\Bigg{[}
-\frac{2}{3}-e^2+\Big{(}1+\frac{3}{2} e^2-\frac{1}{2} S^2\Big{)}P^2 \nonumber \\
&&
+\Big{(}1-\frac{5}{2} e^2-\frac{1}{2} S^2\Big{)} P^2 \cos(2 \varphi^{\ast}-2\ell-2  \omega)
+\frac{7}{2} eP^2\cos(2 \varphi^{\ast}-3\ell-2  \omega)\nonumber\\
&&
-\frac{1}{2} e P^2 \cos(2 \varphi^{\ast}- \ell-2\omega)
+\frac{17}{2} e^2 P^2 \cos(2 \varphi^{\ast}-4  \ell - 2\omega)\nonumber\\
&&
-\Big{(}2-3P^2\Big) e \cos \ell -
\Big(3-\frac{9}{2}  P^2 \Big)e^2 \cos 2\ell\nonumber\\
&&
+QS \Big(\sin \varphi^{\ast}- \sin(\varphi^{\ast}-2\ell - 2\omega )\Big{)} 
+\frac{1}{2} P^2S^2 \Big(\cos 2 \varphi^{\ast}
+ \cos(2\ell+2\omega)\Big)\Bigg{]}\label{U2}
\end{eqnarray} 

This completes the computing of the potential raised by the tidally deformed body on an arbitrary point in space up to the second order in eccentricities and inclinations. 

\section{Tidal waves}

To interpret the terms in $U_2$, we consider a point fixed to the surface of the body. The longitude of this point is $\varphi^{\ast}=\Omega t+\varphi^{\ast}_{0}$, where $\mathbf{\Omega}= \Omega\,\mathbf{\hat{k}}$ is the rotation angular velocity vector of $\tens{m}$ assumed normal to the reference plane and $\varphi^{\ast}_{0}$ is a constant. Each term depending on $\varphi^{\ast}$ in $U_2$ corresponds to a tidal wave traveling on the body with given direction and velocity. All terms contribute, in different ways, to the formation and evolution of the tidal bulge on the body. The terms of $U_2$ may be divided into three groups: 
\begin{enumerate}

\item - {\it Sectorial} components having the form $P^2\cos(2\varphi^{\ast}-\alpha)$, where $\alpha$ is the corresponding phase. The amplitude of these terms is maximum when $P^2$ is maximum, i.e. at the equator, and at the longitudes $\varphi^{\ast}=\alpha/2$ and is decreasing towards the poles. The main term (No. 0 in Table \ref{tabla1}) is a wave with period $\pi/(\Omega-n)$ (i.e. half the synodic rotation period) with two maxima located on the intersection with the equator of the meridian passing through the sub-$\tens{M}$   point and the other on its antipodal. If $n\ll \Omega$, the period is nearly half of the rotation period. The next term (No. 1 in Table \ref{tabla1}) is a wave with period $\pi/(\Omega-1.5n)$ (i.e., larger than half synodic rotation period). The wave has two maxima which are located on two antipodal points on the equator. One of them lies on the meridian of the sub-$\tens{M}$   point when the tide generating body is at the periapsis $\ell=0$ and the other when $\ell=\pi$. The high tide moves, in this case, more slowly than the sub-$\tens{M}$   point. Similar analyses can be done for the other terms showing the argument $2\varphi^{\ast}$. They are often called \emph{semi-diurnal tides} since, on the Earth, they have periods close to 12 hours. They are shown in Table \ref{tabla1}, which also summarizes the interpretation to be given in other cases ($\Omega \ll n$ and $\Omega\simeq n$). 

\item - {\it Zonal} components independent of the longitude $\varphi^{\ast}$. These components of the tidal potential oscillate all over the body with amplitudes depending on the latitude of the points and on the mean longitude of the tide generating body. For example, the component proportional to $(-2+3P^2)\cos \ell $ has relative maximum amplitudes at the equator and the poles, but with inverted phases. The phase inversion occurs at the critical latitude corresponding to $P^2=2/3$ ($\theta^{\ast}=54.7$ degrees). These terms are often called \emph{radial tides} because there is no propagation of a crest around the body. 

\item - {\it Tesseral} components having the form $QS\sin(\varphi^{\ast}-\alpha)$. The amplitude of these terms is maximum when $Q=1$, i.e. at the latitudes $\pm45$ degrees ($\theta^{\ast}=$ 45 and 135 degrees) and at the longitudes $\varphi^{\ast}=\alpha$, with a phase inversion on the equator. They are often called \emph{diurnal tides} since, on the Earth, they have periods close to 24 hours.\\
\end{enumerate}

\begin{table}
\begin{center}
\begin{tabular}{|c|c|c|c|c|}
  \hline
  No. & frequency  &  Type 1 &  Type 2 &  Type 3\\
 && $\Omega \gg n$ & $\Omega \simeq n$ &  $\Omega \ll n$\\ 
  \hline
  \hline
  0 & $2\Omega-2n$   & \small{semi-diurnal} & ${-}$ & \small{semi-annual}\\
  \hline
  1 & $2\Omega-3n$  & \small{semi-diurnal} & \small{monthly} & \small{$3^{rd}$ of annual}\\
  \hline
  2 & $2\Omega-n$  & \small{semi-diurnal} & \small{monthly} & \small{annual}\\
  \hline
  3 & $2\Omega-4n$  & \small{semi-diurnal}  & \small{semi-monthly} & \small{$4^{th}$ of annual}\\
  \hline
  4 & $2\Omega$  & \small{semi-diurnal} & \small{semi-monthly}& \small{``semi-diurnal"}\\
  \hline
  5 & $n$  & \small{monthly} & \small{monthly}& \small{annual}\\
       &      &   (radial)   &   (radial)    &   (radial)\\
  \hline 
  6, 7 & $2n$ & \small{semi-monthly} & \small{semi-monthly}& \small{semi-annual}\\ 
       &      &   (radial)   &   (radial)    &   (radial)\\
\hline
  8 & $\Omega-2n$  &\small{diurnal} & \small{monthly}& \small{semi-annual}\\
\hline
  9 & $\Omega$  & \small{diurnal} & \small{monthly}& \small{``diurnal"}\\
\hline
\end{tabular}
\caption{Tidal potential analyzed term by term. The given tide frequencies and corresponding names refer to how the tidal potential is felt on a given (fixed) point of the body. For the type 2 tides, the paradigm is the Moon, but when the synchronous companion is an exoplanet, the names annual and semi-annual would be more appropriate. }
\label{tabla1}

\end{center}
\end{table}

In table \ref{tabla1}, we summarized the analysis of the tidal waves for three different cases depending on the rotation speed of the deformed body. Type 1 corresponds to a body rotating with angular velocity much larger than the orbital mean motion ($\Omega \gg n$). It is the case of the Earth-Moon system, with the Earth as the deformed body and the Moon as the perturbing one. Type 2 corresponds to \emph{synchronous} or \emph{almost synchronous} motions and, again, the Earth-Moon system serves as an example, but now the Moon is the deformed body and the Earth is generating the tide. Looking at Table \ref{tabla1}, we see that synchronization gives rise to terms whose period is related to the rotation period of the companion, and they are called monthly, semi-monthly, etc. The names come from the tidal action of the Earth on the Moon (for this reason, the semi-monthly tide is often called {\it fortnightly}). Since $\Omega \simeq n$, we could also use the names diurnal, semi-diurnal, etc. (referring to the companion's rotation). Type 3 corresponds to a slow rotating body ($\Omega \ll n$); it is the case of the tides on a typical main sequence star due to a close-in planet (hot Jupiter). Using names similar to those used in the other cases and taking into account that the main period is the planet's  orbital period (or ``year"), we will call them, respectively, annual, semi-annual, tierce-annual and so on\footnote{We have, however, to keep in mind that these ``years" are very short. For instance, the orbital period of OGLE-TR-56b, one of the shortest known, is only 1.21 days. In type 3 tides, ``diurnal" is much slower than ''annual".}. 

It is worth emphasizing that the given tidal frequencies and corresponding names refer to how the tidal potential is felt on a given (fixed) point of the body. The propagation of the tidal wave in the body must be analyzed separately. For instance, on the Earth, the tidal bulges of both diurnal and semi-diurnal tides circulate around the Earth with the synodic rotation speed. The names and frequencies given in Table \ref{tabla1} refer rather to the shape of the tidal wave.
   
Table \ref{tabla1} is limited to the tidal components appearing in the given expansion of $U_2$. When higher order terms are considered, many other frequencies appear. 

\section{The tidal phase lags}\label{lags}

In the previous sections, we have considered that $\tens{m}$ is a perfect inviscid fluid that reaches the equilibrium figure instantaneously under the attraction of $\tens{M}$. However, in a real body, the viscosity introduces a delay between the tidal action and the corresponding response. 

The main characteristic of Darwin's theory is to consider the potential $U_2$ as a composition of periodic terms with different frequencies and to introduce in each periodic term a delay in the form of a \emph{lag angle} (Darwin, 1880). $U_2$ is then expanded to first order in the lags. The trigonometric functions are expanded in the following way:
\begin{eqnarray}\label{epsilon}
\cos(\Phi_{i}-\ep_{i})&\simeq&\cos\Phi_{i}+\ep_{i}\sin\Phi_{i}\\
\sin(\Phi_{i}-\ep_{i})&\simeq&\sin\Phi_{i}-\ep_{i}\cos\Phi_{i}
\end{eqnarray}
where $\Phi_i$ is a generic time-dependent argument.

This is not the only way of introducing the lags. We may mention the theories of MacDonald (1964) and Mignard (1979, 1980) as paradigms of different approaches. In both cases, the \emph{tidal lag} is associated with the displacement of the tidal bulge dragged by the rotation of the body. In MacDonald's theory, the tidal lag is a constant and, in Mignard's theory, the tidal lag is proportional to the relative (synodic) rotation speed. 
Eggleton et al. (1998) use a different physical approach which, however, leads to the same tidal force given by Mignard (1979).
When these theories are used to study Earth's bodily tides, their results do not differ essentially from those obtained by Darwin. Indeed, on Earth, the tidal effects are dominated by semi-diurnal tides, and the different approaches address more or less the same problem. Difficulties arise in the case of synchronous and quasi-synchronous rotation. The classical theories show that, in absence of non-tidal torques, no synchronization is possible if the orbital eccentricity is not damped to zero. The tidal torque on the deformed body vanishes for a rotation velocity slightly larger than the orbital mean motion. A simple physical reasoning shows that this result is expected. Indeed, the tidal torque is inversely proportional to the sixth power of the radius vector (see eqn. \ref{torques}). When the orbit is eccentric, the torque will be much greater at the periapsis than at the apoapsis, and, thus, the average will correspond to a rotational angular velocity exceeding the orbital mean motion. Therefore, 
the rotation velocity in the stationary solution is larger than the orbital mean motion. Linear theories with a constant time lag give for the stationary rotation velocity: 
\beq\Omega=n(1+6e^2)\endeq
(see eqn. \ref{Omegadot}). 
In the case of Titan, this result corresponds to $\Omega=n(1 + 5 \times 10^{-3})$, that is, to a synodic rotation period of about 8.5 yrs, which would be observable notwithstanding the difficulties in the identification of features in the surface of Titan. Several attempts were done to modify Mignard's and similar theories by introducing phase lags depending nonlinearly on the wave frequencies (see e.g. Sears et al. 1993). 

The question of the law that should be used to relate tidal lags and frequencies is a controversial one. Some results obtained from the study of the Moon (Williams et al. 2006) indicate that the variation, for a large range of frequencies, is small: the quality factor\footnote{
The quality factor is usually introduced through its relationship to the geometric lag angle $\Delta$: $ Q \sim 1/2\Delta$ (see MacDonald, 1964; Efroimsky and Lainey, 2007); the corresponding relationship when the phase lag of the semi-diurnal tide is used instead of the geometric phase angle is $Q \sim 1/\ep_0$. This definition needs to be amended when $\ep_0$ becomes small. See sect. \ref{secQ}.} 
$Q$ increases from 30 for one month to 34 for one year. This corresponds to a power law with an exponent $\sim 0.04$. On the other hand, Efroimsky and Lainey (2007) collected geophysical data which indicate, in the frequency range of our concern, an inverse law: the dependence of lag on the frequency follows a power law with a negative exponent in the range $(-0.4,-0.2)$.  

Darwin's theory introduces naturally different lags for different terms without the need of introducing a priori a particular law (notwithstanding the fact that Darwin himself used a linear law). We will just assume that equal frequencies correspond to equal lags and that the lag vanishes when the frequency tends to zero. We also pay attention to the fact that some frequencies in Table \ref{tabla1} become negative when $\Omega \simeq n$ or $\Omega \ll n$ (Types 2 and 3 tides). The rule is that the actual phase of the terms in the dynamic equilibrium tide lags behind the corresponding phase in the ``static" case.
Thus, in type 2 tidal interaction, the lags may be such that $\ep_{1}\sim -\ep_{2} < 0$ and $\ep_{8}\sim -\ep_{9} < 0$. In the same way, in type 3, we have $\ep_i < 0$ for all subscripts corresponding to negative frequencies ($i=0,1,2,3,8$).

\section{The dynamic equilibrium tide}

In addition to the phase lag of the tidal waves, we may assume that the body does not reach total deformation and thus substitute, in the coefficients, the factor $k_f$ by dynamical counterparts $k_i$ which may be assumed to depend on the frequencies of the corresponding tidal waves. However, in first-order theories, every $k_i$ appears in the tidal potential multiplied by the corresponding lag $\ep_i$. We will put in factor one value $k_d$ (called \emph{dynamic Love number}) for the main tidal wave (which, for the Type 1 of tidal interaction may be taken as the semi-diurnal Love number $k$) and merge the others with the corresponding phase lag. The lags $\ep_i$ and the response factors $k_i$ are very distinct physical quantities but, as far as only the potential due to $\tens{m}$ is considered, they do not need to be considered separately. We may keep in mind that when physical interpretations are sought, it is easy to separate their contributions. To avoid any misinterpretation, we introduce the modified lags  $\epp_j=(k_j/k_d)\ep_j$ and use them instead of $\ep_j$.

Thus, instead of $U_2$, we have

\begin{equation}\label{zero+lag}
U_{2}=U_2^0+U_{\rm lag}
\end{equation}
where $U_2^0$ is given by eqn. (\ref{u}) and
\begin{eqnarray}\label{ulag}
U_{\rm lag}&=&-\frac{3k_dGMR^{5}}{8a^3r^{\ast 3}}\Bigg{[}
P^2  \epp_0  \Big(2 - 5 e^2  - S^2\Big)\sin(2 \varphi^{\ast}-2 \ell-2\omega)\nonumber\\&&
+ e P^2    \Big(7\epp_1\sin(2 \varphi^{\ast} -3 \ell - 2\omega  )
- \epp_2 \sin(2 \varphi^{\ast}- \ell -2\omega)\Big)\nonumber\\&&
+{17} e^2  P^2  \epp_3 \sin(2 \varphi^{\ast} -4 \ell -2\omega)
+  P^2 S^2  \epp_4 \sin 2 \varphi^{\ast}\nonumber\\&&
- e \epp_5  (4-6 P^2)\sin \ell 
-{3} e^2  \epp_6 (2-3 P^2 )\sin 2\ell\nonumber\\&& 
+  P^2 S^2 \epp_7 \sin(2\ell + 2\omega)
+2QS \Big(\epp_8 \cos( \varphi^{\ast} - 2 \ell -2\omega ) 
- \epp_9 \cos\varphi^{\ast} \Big) \Bigg].
\end{eqnarray}
We recall that, in classical theories, in which tides on the Earth are the only ones considered, the lags are introduced as $2\ep$ because the main terms correspond to semi-diurnal tides (see Jeffreys, 1961). In this report, all lags are introduced as $\ep_i$, in a unified way.

\section{Tidal forces acting on the tide generating body}\label{forces}

The perturbing force acting on a point of mass $M^{\ast}$ placed in $\mathbf{r}^{\ast}\equiv (r^{\ast},\theta^{\ast},\varphi^{\ast})$, due to the disturbing potential $U_2$ is given by 

\begin{eqnarray}\label{grad}
\mathbf{F}=-M^{\ast}{\rm grad}_{{\mathbf r}^\ast} U_{2}=&\underbrace{-M^{\ast}\frac{\partial U_{2}}{\partial r^{\ast}}}&\mathbf{\hat{r^{\ast}}}\underbrace{-\frac{M^{\ast}}{r^{\ast}}\frac{\partial U_{2}}{\partial \theta^{\ast}}}\mathbf{\hat{\theta^{\ast}}}\underbrace{-\frac{M^{\ast}}{r^{\ast}\sin\theta^{\ast}}\frac{\partial U_{2}}{\partial\varphi^{\ast}}}\mathbf{\hat{\varphi^{\ast}}}\\
&F_{1}&\qquad F_{2}\qquad\qquad\qquad F_{3}
\end{eqnarray}
where $(\hat{r^{\ast}},\hat{\theta^{\ast}},\hat{\varphi^{\ast}})$ form, at $\mathbf{r}^{\ast}$, a right-handed orthogonal set of unit vectors in the positive direction of the increments. ($\hat{\varphi^{\ast}}=\hat{r^{\ast}}\times\hat{\theta^{\ast}}$.) The total force is decomposed into three orthogonal components. Knowing that ${\rm grad}_{{\mathbf r}^\ast} U_{2}={\rm grad}_{{\mathbf r}^\ast} U_2^0+{\rm grad}_{{\mathbf r}^\ast} U_{lag}$, we may calculate the forces separately for each part of the disturbing potential and add them later.
These forces are given in Appendix C of the astro-ph version of the report. They are important intermediaries in the calculations, but have no importance per se.

To obtain the forces on the tide generating body $\tens{M}$, it is enough to make the identification of $(M^{\ast},r^{\ast},\theta^{\ast},\varphi^{\ast})$ and $(M,r,\theta,\varphi)$. For sake of completeness, we note that this identification is done after the calculation of the gradient of $U_2$. 

The forces derived from $U_2^0$ are
\begin{equation}\label{f1u2}
F_{1}^{U}=-\frac{3k_fGM^2R^{5}}{r^{7}},
\hspace{1cm} F_{2}^{U}=F_{3}^{U}=0.
\end{equation}
Only the radial force survives in the absence of lags, as expected from the symmetry of the resulting configuration (see figure \ref{esferoide}).

Making the identification in the forces arising from $U_{\rm lag}$ and introducing the expressions given by eqns. (\ref{cosfi})--(\ref{raster}), we obtain the force components 

\begin{eqnarray}\label{f1ueva}
F_{1}(\textbf{r})&=&\frac{3k_dGM^2R^{5}}{8a^7}\Bigg{[}
-3e(8\epp_0-7\epp_1-\epp_2+2\epp_5)\sin\ell\\&&-3e^2(21\epp_0-4\epp_2-17\epp_3+4\epp_5+3\epp_6)\sin 2\ell
+3S^2(\epp_0-\epp_4-\epp_7-2\epp_8+2\epp_9)\sin(2\ell+2\omega)\Bigg{]}\nonumber
\end{eqnarray}
\begin{eqnarray}\label{f2ueva}
F_{2}(\textbf{r})&=&\frac{3k_dGM^2R^{5}}{8a^7}\Bigg{[}
eS (+8\epp_0-7\epp_1-\epp_2+6\epp_5
-12\epp_8+4\epp_9+6\epp_{15}+2\epp_{16})\cos \omega \\&&
-eS(8\epp_0-7\epp_1-\epp_2+6\epp_5+4\epp_8-12\epp_9-6\epp_{14}+14\epp_{17})
\cos(2\ell+\omega)
-4S (\epp_8-\epp_9)\cos(\ell+\omega)\Bigg{]}\nonumber
\end{eqnarray}
\begin{eqnarray}\label{f3ueva}
F_{3}(\textbf{r})&=&\frac{3k_dGM^2R^{5}}{8a^7}\Bigg{[}(4-14e^2-3S^2)\epp_0+56e^2\epp_1+2S^2(\epp_8+\epp_9)
+e(16\epp_0+14\epp_1-2\epp_2)\cos\ell\nonumber\\&&
+e^2(44\epp_0-8\epp_2+34\epp_3)\cos 2\ell
+S^2(\epp_0+2\epp_4-2\epp_8-2\epp_9)\cos(2\ell+2\omega) \Bigg{]}.
\end{eqnarray}
The terms with the lags $\epp_{14},\epp_{15},\epp_{16},\epp_{17}$ introduced in $F_2$ ({\it cf.} Errata in press) come from $\delta_3U_{\rm lag}$ (see Appendix D). When we assume the lags proportional to the frequencies, the forces given by eqns. (\ref{f1ueva} -- \ref{f3ueva}) are equivalent to the second-degree expansion of the force given by Mignard (1979).

The average values of the $F_{i}$ can be easily found from the above equations. However, in the used spherical coordinates, the unit vectors are continuously changing and averages are of limited interest. We nevertheless note that the average radial component of the tidal force acting on $\tens{M}$ is zero and that, when $I=0$, the component $F_2$ vanishes.

\section{The tidal torque}

Because of the delay in the response to tide raising forces, the tidal bulge is not aligned with $\mathbf{r}$. 
This causes the raising of a tidal torque 
\begin{equation}\label{torque}
\cal{M}=\mathbf{r}\times\mathbf{F}
\end{equation}
or, since $\mathbf{r}\equiv(r,0,0)$, 
\begin{equation}\label{torque2}
{\cal{M}}\equiv(0,M_2,M_3)=-rF_{3}\hat{\theta}^{\ast}+rF_{2}\hat{\varphi}^{\ast}.
\end{equation}

\begin{figure}[t]
\centerline{\hbox{
\includegraphics[height=3.5cm,clip=]{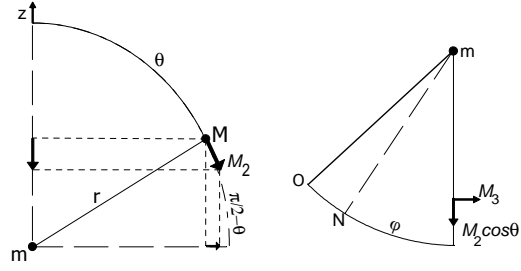}}}
\caption{Projection of the torque components on the meridian plane of $\tens{M}$ (left) and on the reference plane (right)}
\label{eplot2}    
\end{figure}

We do not write explicitly the expressions for the two non-zero components of $\cal{M}$, as they are simply products of the force components by $r$. On the other hand, in the applications, we need the average torque, which needs to be calculated in a fixed reference frame. We thus transform $\cal{M}$ to obtain a new decomposition. This is easily done since $M_1=0$ and $M_3$ is orthogonal to the meridian plane of $\tens{M}$ (see fig. \ref{eplot2}). If, for simplicity, we adopt here a system whose $x$-axis is oriented towards the ascending node $N$, we obtain
\begin{eqnarray}
M_x&\speq&M_2\cos\theta \cos\varphi - M_3 \sin\varphi\nonumber\\
M_y&\speq&M_2\cos\theta \sin\varphi + M_3 \cos\varphi\\
M_z&\speq&-M_2\sin\theta . \nonumber
\end{eqnarray} 

The averages of $M_x,M_y,M_z$ over one orbital period are 

\begin{eqnarray}
< M_x >&\speq& 0\nonumber\\
< M_y >&\speq& -\frac{3k_dGM^2R^{5}}{4a^6}S(\epp_0+\epp_8-\epp_9) \label{torques} \\
< M_z >&\speq& \frac{3k_dGM^2R^{5}}{8a^6}\Big(4\epp_0 
+e^2(-20\epp_0+49\epp_1+\epp_2)+2S^2(-2\epp_0+\epp_8+\epp_9)\Big). \nonumber
\end{eqnarray} 

The above results deserve some comments: (1) The average torque is perpendicular to the line of nodes (perpendicular to the orbital plane if $I=0$); (2) The results in a system whose axis is directed to a fixed point O (see fig. \ref{eplot1}) are easily obtained rotating the results by an angle $\widehat{\rm {ON}}$ (longitude of the ascending node) around the $z$-axis; 

\section{The rotational angular momentum}

The angular momentum conservation is very stringent and cannot be used without the simultaneous analysis of all involved forces. For instance, the possible non-spherical shape of the bodies and the tidal friction in both bodies (see Section \ref{centralbody}) will move the planes considered in Fig. \ref{coord}. In this section, we consider only the interaction between the orbit and the tidally deformed body. If we neglect all additional factors mentioned above, the change of angular momentum in the orbit may be compensated by a change in the angular momentum of the tidally deformed body of opposite direction. We may have $\dot{\cal L}_{\rm rot} + \dot{\cal L}= 0$; but $\dot{\cal L} = {\cal M}$. Hence $\dot{\cal L}_{\rm rot} = -{\cal M}$. 
A conceptual difficulty in this calculation is the fact that we are using the equator as reference plane and it is no longer inertial if moving. To overcome this difficulty the simplest way is to consider the equator at a given fixed time inside the considered interval as reference plane. Alternatively, we could add the centrifugal and Coriolis forces acting on the bodies due to the motion of the reference plane. We adopt here the first of these two approaches. The second approach is discussed in Appendix E of the astro-ph version of this report.

The rotational angular momentum of $\tens{m}$ is given by ${\cal L}_{\rm rot}\simeq C\Omega\, \hat{\mathbf{k}}$ where $C$ is the moment of inertia of $\tens{m}$ with respect to the principal axis $c$ (see fig. \ref{esferoide}), $\Omega$ is the angular velocity of rotation of $\tens{m}$ and $\hat{\mathbf{k}}$ is a unit vector along the principal axis $c$.      

\begin{figure}[t]
\centerline{\hbox{
\includegraphics[height=3.5cm,clip=]{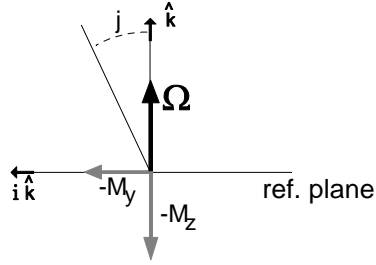}}}
\caption
{Meridian plane normal to the nodal line. Components of the torque on the tidally deformed body. $\hat{\mathbf{k}}$ and $\dot{\imath}\hat{\mathbf{k}}$ are unit vectors. $\mathbf{\Omega}$ is the angular velocity vector}
\label{eplot5}    
\end{figure}

Since the average torque ${\cal M}$ is normal to the line of nodes, we may decompose the equation $\dot{\cal L}_{\rm rot}=-{\cal M}$ into two parts and study each part separately.  In the meridian plane normal to the nodal line, the time derivative of the angular velocity vector is $\frac{d}{dt}{\Omega}\,{\hat{\mathbf{k}}}=\dot{\Omega}\, \hat{\mathbf{k}} + \Omega \dot{J}\, \dot{\imath}\hat{\mathbf{k}}$ where $\dot{J}$ is the variation of the inclination due to the variation of the unit vector $\hat{\mathbf{k}}$ (a rotation). We introduced a new letter ($\dot{J}$) to distinguish between this quantity and the variation of the inclination due to the motion of the orbital plane ($\dot{I}$). From the given equation, we obtain $C\dot{\Omega}=-M_z$ and $C\Omega \dot{J}= -M_{y}$.

Then, taking the averages,

\beq
<\dot{\Omega}> \speq - \frac{3k_dGM^2R^{5}}{8Ca^6}\Big[4\epp_0 
+e^2(-20\epp_0+49\epp_1+\epp_2)+2S^2(-2\epp_0+\epp_8+\epp_9)\Big] \label{dotOmega}
\endeq\beq
<\dot{J}> \speq \frac{3k_dGM^2R^{5}}{4C\Omega a^6}S(\epp_0+\epp_8-\epp_9). 
\label{dotJ}\endeq

\section{Rotation of Close-in Companions}\label{close-in}

Let us now assume that the body $\tens{m}$ is a satellite or exoplanet orbiting close to a major primary body. Let us also assume that it has no other significant deformation besides the tidal one due to the central body. 
Eqn. (\ref{dotOmega}) shows that, in the first approximation, $<\dot{\Omega}> \,\propto \epp_0$, indicating that the system will smoothly evolve towards synchronization.
However, the final state is not necessarily synchronous.

Close to synchronization, we have a type 2 tide and several lags correspond to monthly tides (see table \ref{tabla1}). 
We assume that they are equal, just taking as negative those corresponding to negative frequencies: 
\beq
\epp_2 \simeq -\epp_1 \simeq \epp_9 \simeq -\epp_8>0.
\label{type2cond}\endeq
Hence,
\beq
<\dot{\Omega}> \speq - \frac{3k_dGM^2R^{5}}{2Ca^6}\Big(\epp_0 
-e^2(5\epp_0+12\epp_2)-S^2\epp_0\Big) \label{dotOmegaclose}
\endeq\beq
<\dot{J}> \speq \frac{3k_dSGM^2R^{5}}{4C\Omega a^6}(\epp_0-2\epp_2). 
\endeq

\subsection{Stationary rotation}

We say that the system reaches a state of stationary rotation when the average angular acceleration (or the average tidal torque $<M_z>$) vanishes.
Solving the equation
$<\dot{\Omega}> \speq 0,$
we obtain, at the second order in $e,I$, 
\begin{equation}\label{ep0}
\epp_0\speq 12 e^2 \epp_2.
\end{equation}      
In linear theories with constant time lag (see Section \ref{linear}), each $\epp_i$ is assumed proportional to the frequency of the corresponding tidal wave. When these $\epp_i$ are substituted in the previous equation, we obtain the well-known result\footnote{When the tidal phase lag is assumed to be frequency independent (MacDonald theory), the resulting stationary velocity is $\Omega_{\rm stat}\speq n(1+9.5\,e^2)$ (Goldreich, 1966).} 
\beq
\Omega_{\rm stat}\speq n(1+6e^2),
\label{Omegadot}\endeq
showing that, when $e \ne 0$, 
the rotation stabilizes at a slightly super-synchronous value. 
Nonlinear theories give similar results only modifying the numerical factor multiplying $e^2$ (Sears et al., 1993).

The inclination creates an effect tending to sub-synchronize the stationary rotation, but it is of order ${\cal O}(S^4)$ and therefore beyond the order of approximation adopted in this report (see Levrard, 2008).

The second equation resulting from the angular momentum conservation gives
\begin{equation}
<\dot{J}> \speq -\frac{3k_dSGM^2R^{5}}{2C\Omega a^6}\epp_2(1-6e^2).
\endeq
where one mixed third order term appearing due to the substitution of $\epp_0$ by its value was kept to show one of the few instances in which terms of this kind affect the result.  
 
The part of the equation corresponding to the vanishing of the torque along the nodal line indicates that no precession of the nodes occurs when only these forces are considered. 

\subsection{Spin-orbit synchronization}

The spin-orbit synchronization condition is given by $\Omega = n$.
When $\epp_0=0$, eqn. (\ref{dotOmegaclose}) gives
\beq
<\dot{\Omega}> \speq  \frac{18k_dGM^2R^{5}}{Ca^6} e^2 \epp_2,
\endeq
which cannot vanish if $e \ne 0$ and is thus inconsistent with a synchronization.
It is worth noting that $S$ does not affect the above condition (at least when terms of the fourth order in $e,I$ are neglected).

\subsection{Synchronous asymmetric companions}\label{secU22}

The above conclusions were derived from the assumption that $\tens{m}$ has no other significant deformation besides the tidal one. If the close-in companion has a permanent (solid-like) equatorial ellipticity, different results follow. Indeed, in such case, we have to add the contribution of the equator asymmetry to $U_{\rm lag}$, e.g.
\begin{equation}
{U}_{22}=\frac{GmR^2}{r^3}J_{22}P_{22}(\cos\theta)\cos 2(\varphi-\varphi_{22})
\label{U22}\endeq
where $P_{22}$ is an associated Legendre function. $J_{22}>0$ and $\varphi_{22}$ are the two parameters characterizing the  asymmetry of the gravitational field\footnote{The sign adopted for $U_{22}$ is the one more currently used (see Beutler, 2005).  $\varphi_{22}$ is the longitude of one point on the shortest of the two equatorial axes.}. According to Goldreich (1966), synchronous rotation will result when $J_{22}$ is larger than a critical value (see below). 
In such a case, the body will end up with spin-orbit synchronization and,
as a consequence, we have, 
\beq \ep_0=0,\endeq 
instead of eqn. (\ref{ep0}).

In this case, the equations obtained in the previous section can no longer be used. The synchronous motion is also a stationary solution. However it is a stationary solution of one system in which the companion's rotation is under the simultaneous action of two torques: the tidal torque and the torque due to the asymmetry of the body, which allows the motion to become synchronous. 

Since the tidal friction in a synchronous companion moving in an eccentric orbit tends to accelerate its motion, the body will be displaced forward but, being asymmetric, this displacement will create a torque in the contrary direction which will compensate the tidal torque. The average torque due to the displacement by an angle $\delta$ forward is
\beq
<(M_{22})_z>\speq\frac{6GMmR^2J_{22}\sin 2\delta}{a^3}\,\big(1-\frac{5}{2}\,e^2-\frac{1}{2}\,S^2\big).
\endeq
The condition $<\dot{\Omega}>=0$ means, now, $<M_z+(M_{22})_z>=0$, which allows the average offset angle $\delta$ to be determined.  Using the condition $\epp_0=0$, there follows, at the order of approximation adopted in this paper,
\beq
J_{22}\sin 2\delta \simeq \frac{3MR^3}{ma^3}k_d\epp_2e^2.\label{J22crit}
\endeq
If we know the maximum value that $\delta$ can reach without disturbing the capture into the 1:1 spin-orbit resonance, we may obtain the minimum value of $J_{22}$ necessary to capture into the 1:1 resonance.\footnote{When the tidal phase lag is assumed to be frequency independent (MacDonald theory), the resulting critical value of $J_{22}$ is proportional to $e^4$ (Goldreich, 1966). However, with the averaged equations of this section, issued from Darwin's theory, it appears proportional to $e^2$.}
The numerical exploration of some examples has shown here a large influence of the periodic terms which may increase the critical value of $J_{22}$ by several orders of magnitude.  

The torque due to $U_{22}$ also contributes to the motion of the equatorial plane of the synchronous companion. The $y$-component of the momentum is, in this case, proportional to $SJ_{22}\sin 2\delta$ (that is, to $Se^2\epp_2$) and is negligible. The $x$-component is 
\beq
<(M_{22})_x>\speq -\frac{3GMmR^2J_{22}S}{a^3}\cos 2\delta.
\endeq
This term will give rise to an acceleration in the plane defined by the rotation axis and the nodal line. 
This is the classical picture in which the rotation vector describes a precession cone and is the counterpart, in the body, of the precession of the nodal line given by eqn. (\ref{Node22}). (The total angular momentum of the system is invariant to the action of $U_{22}$.) The approach used here is not adequate to study it. Moreover, the dependence of $(M_{22})_x$ on $\delta$ is of higher order and, in the first approximation, $(M_{22})_x$ can be written in a form independent of $\delta$:
\beq
<(M_{22})_x> \,\simeq -\frac{3GMmR^2J_{22}S}{a^3}.
\endeq
$<(M_{22})_x>$ is, thus, rather due to the figure of the companion trapped into the 1:1 spin-orbit resonance and exist even when the tide is neglected (i.e. when $\delta \simeq	 0$). The resulting precession is not an indirect tidal perturbation.

It is worth mentioning that if, instead of ${U}_{22}$, we introduce 
\begin{equation}
{U}_{31}=\frac{GmR^3}{r^4}J_{31}P_{31}(\cos\theta)\cos (\varphi-\varphi_{31})
\endeq
where $P_{31}$ is an associated Legendre function and $J_{31}$ and $\varphi_{31}$ the two parameters characterizing the deformation of the equator (see Beutler, 2005), the results are very similar. We do not reproduce details, but just say that in this case, instead of eqn. (\ref{J22crit}), we obtain
\beq
J_{31}\sin \delta' \simeq \frac{12MR^2}{ma^2}k_d\epp_2e^2.
\endeq
If necessary, it is easy to consider $U_{22}$ and $U_{31}$ simultaneously.

\section{The Work done by the Tidal Forces}\label{work}

The work done by the tidal forces in a displacement $d\mathbf{s}$ is given by $dW=\mathbf{F}\cdot d\mathbf{s}$, or $\dot{W}=\mathbf{F}\cdot \mathbf{v}$ where $\mathbf{v}$ is the velocity vector. This calculation is elementary. The only necessary precaution is to take into account that the usual expressions for the radial and transverse components of the velocity,
\begin{eqnarray}
v_{\rm R}&\speq&an\Big(e\sin\ell+e^2\sin 2\ell\Big)\nonumber\\
v_{\rm T}&\speq&an\Big(1+e\cos\ell+e^2\cos 2\ell-\frac{1}{2}e^2\Big),
\end{eqnarray}
are given in a reference system whose plane is the orbit plane. The radial components in this system and in the spherical coordinates used in this study are the same. However the transverse component lying on the orbital plane needs to be decomposed along the axis of the local spherical coordinates frame: $v_2=-v_{T}\sin\beta$ and $v_3=v_{T}\cos\beta$  (see Fig. \ref{eplot3}). 
From the triangle shown in the figure, we obtain (by sine and cosine laws):
$\cos\beta=\sin\varphi/\sin(v+\omega)$ and
$\sin\beta=S \cos\varphi$
\begin{figure}[h]
\centerline{\hbox{
\includegraphics[height=3.5cm,clip=]{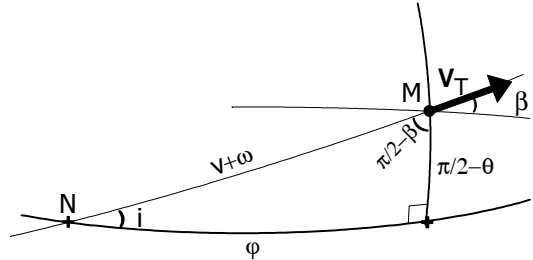}}}
\caption
{Transverse velocity at $\tens{M}$ and circle tangent to the parallel.}
\label{eplot3}    
\end{figure}

The work is then easily calculated by the given scalar product. The result, averaged over one orbit is
\beq
<\dot{W}>_{\rm tide} \speq \frac{3nk_dGM^2R^5}{8a^6}\Big[ 4\epp_0+e^2(-20\epp_0+\frac{147}{2}\epp_1+\frac{1}{2}\epp_2-3\epp_5)-4S^2(\epp_0-\epp_8)\Big].\label{work-av}
\endeq

\subsection{Energy Dissipation}

The energy release may be known from the fact that the total energy variation of the system must be equal to zero. Besides the orbital energy, whose variation is given above, we have the rotational energy of the deformed body and the thermal energy dissipated in the body. The balance equation allows us to calculate the energy dissipated in the deformed body. 

For that sake, we have to know the work done by the tidal torque acting on the body 
(which is the opposite of the torque acting on $\tens{M}$),
\begin{eqnarray}
<\dot{W}>_{\rm rot} &\speq &<-{\cal{M}}\cdot\mathbf{\Omega}> \speq C\Omega<\dot{\Omega}> \\ \nonumber
&\speq &
 -\frac{3\Omega k_dGM^2R^{5}}{8a^6}\Big[4\epp_0 
+e^2(-20\epp_0+49\epp_1+\epp_2)+2S^2(-2\epp_0+\epp_8+\epp_9)\Big]
\label{work-rot} 
\end{eqnarray}
If the rotation of the considered body is relatively fast (i.e. $\,\Omega > n,\, \epp_0>0$), rotational energy is transferred from its rotation to the orbital motion while, if it is slow (i.e. $\,\Omega < n,\, \epp_0<0$), orbital energy is lost and partly used to accelerate its rotation. In both cases, $<\dot{W}>_{\rm tide+rot}\, < 0$ and the mechanical energy lost by the system must be released inside the body with the average rate \mbox{$<\dot{E}_{\rm th} > \simeq - <\dot{W}>_{\rm tide+rot} $}. 

\section{Energy dissipation in close-in companions}

We consider here the case of companions having reached a final rotation with $\Omega \simeq n$. In the resulting type 2 tide, we may take into account the condition stated by eqn. (\ref{type2cond}): $\epp_2 \simeq -\epp_1 \simeq \epp_9 \simeq -\epp_8>0$. 
We assume also that the lag $\epp_5$ of the radial tide is equal to the lag $\epp_2$ of other tidal waves with the same period. Hence
\beq
<\dot{W}>_{\rm tide} \speq \frac{3nk_dGM^2R^5}{2a^6}\Big[
 \epp_0-e^2(5\epp_0+19 \epp_2)-S^2(\epp_0+\epp_2)\Big]
\endeq
and
$<\dot{W}>_{\rm rot} \simeq 0$ (since $\dot{\Omega} \simeq 0$).

\subsection{Stationary rotation}

If the only torque acting on the body in stationary rotation is the tidal torque, we have 
$\epp_0 = 12e^2\epp_2$ and the above equation becomes 
\beq
<\dot{W}>_{\rm tide}\,
\simeq -\frac{3nk_dGM^2R^5}{2a^6}\,(7 e^2 + S^2)\epp_2.
\label{dotWstat}\endeq

In this case, the thermal energy released inside the companion is $\dot{E}_{\rm th} \simeq - \dot{W}_{\rm tide}$ (they are equal when we can neglect the changes in the equilibrium rotation speed due to the variation of the mean motion, which is of the order of $(R/a)^2$.) 
Therefore, the dissipation is proportional to the lag $\ep_2$ of the monthly tide (the tide harmonic whose period equals the orbital period).

It is worth emphasizing that this result does not depend on any hypothesis linking lags to frequencies (the only assumption is that equal frequencies lead to equal lags).

\subsection{Synchronous asymmetric companions}\label{secW22}

Proceeding as above and noting that $\epp_0=0$, we obtain, in this case, the result
\beq
<\dot{W}>_{\rm tide} \speq -\frac{3nk_dGM^2R^5}{2a^6}\,(19 e^2+S^2)\epp_2.
\label{dotWsync}
\endeq
However, if $e\ne 0$, a synchronous rotation cannot exist without additional forces acting in the system, which will also do a work. As they counteract the tidal forces, the work is expected to have a different sign. From (\ref{U22}), we indeed obtain
\beq
<\dot{W}>_{22} \speq \frac{6GMmR^2nJ_{22}\sin 2\delta}{a^3}\,
\big(1-\frac{5}{2}\,e^2-\frac{1}{2}\,S^2 \big) 
\simeq \frac{18nk_dGM^2R^5}{a^6}\,e^2\epp_2,
\endeq
where the rightmost expression was obtained introducing the value of $J_{22}\sin 2\delta$ obtained in section \ref{secU22}.

When this term is added to $<\dot{W}>_{\rm tide}$, the result is 
\beq
<\dot{W}>\, 
\simeq -\frac{3nk_dGM^2R^5}{2a^6}\,(7 e^2 + S^2)\epp_2,
\label{dotWtot}\endeq
which is formally equal to that obtained for the stationary case. 
It may be compared to that given by Wisdom (2004) and Winn and Holman (2005): 
\beq
<\dot{E}_{\rm th}> \speq \frac{9nhGM^2R^5}{10Qa^6}\,(7e^2 + S^2).
\label{WisW}\endeq
This result only differs from the given one for using the relationship $h=5k_d/3$ between Love numbers\footnote{$h=5k/3$ in a homogeneous companion. See Munk and MacDonald (1960).}  $h$ and $k_d$ and by adopting $Q=1/\ep_2$. The difference of sign comes from the fact that $E_{\rm th}$ is the energy released in the body.

The same value for the energy dissipation was obtained by Segatz et al (1988) and Wisdom (2008) through the direct calculation of the energy released inside the companion. It is worth noting that the same result can be obtained without making explicit use of $<M_{22}>$. It is sufficient to use the fact that the counteracting torque must be equal (and opposite) to the mean torque accelerating the rotation in the synchronous case and obtain the work done by multiplying the torque by the angular velocity of the body
(see Levrard, 2008).

\subsection{On dissipation and lags}\label{secQ}

Lags and dissipation are different aspects of the same phenomenon.
Viscosity driven tidal friction is directly responsible for the production of heat inside the body and, at the same time, by delaying the response of the body to external forces. 
From the mathematical point of view, we may fix one of them and obtain the other. 
In this paper, we introduced the lags and computed the thermal energy released.

The results of the previous sections show how lag and dissipation are interrelated.
Equations (\ref{work-av}) and (\ref{work-rot}) show that, before synchronization, the loss of orbital and rotational energies due to dissipation in the companion are, in the first approximation, proportional to the lag $\ep_0$ (through $\epp_0$). $\ep_0$ is the lag of the tide whose period is half the synodic rotation period (semi-diurnal tide). It is half the geometric lag (delay of the high tide with respect to the sub-$\tens{M}$   point). In terms of $\ep_0$, the quality factor $Q$ is then usually defined as $Q=1/\ep_0$ and is singular when $\ep_0 \rightarrow 0$. This singularity needs to be explained. In fact, the parameter that measures dissipation is not $Q$ but $1/Q$ (see Munk and MacDonald, 1960, chap. 3), thus the real problem is not the division by zero, but the fact that the given formula leads to $1/Q \rightarrow 0$.  

However, looking at eqns.(\ref{work-av}) and (\ref{work-rot}), we see that the right-hand sides do not vanish when $\ep_0 \rightarrow 0$. Some terms indeed vanish, but others do not and the latter will be the leading terms in the resulting expressions of $<\dot{W}>_{\rm tide}$ and $<\dot{W}>_{\rm rot}$. 

The comparison of eqns. (\ref{dotWtot}) and (\ref{WisW}) has already shown that the usually adopted quality factor is, in this case, $Q=1/\ep_2$. 
It is no longer the inverse of the lag of the semi-diurnal tide, but the inverse of the lag of the tide whose period is the orbital period (monthly tide). 

\section{The variation in semi-major axis and mean motion}\label{sec12}

From $E=-GmM/2a$ and Kepler's law, we obtain, 
\beq
<\dot{n}> \speq -\frac{3n}{2a}<\dot{a}> \speq -\frac{9n^2k_dMR^5}{8ma^5}\Big[
4\epp_0-e^2(20\epp_0-\frac{147}{2}\epp_1-\frac{1}{2}\epp_2+3\epp_5)-4S^2(\epp_0-\epp_8)\Big]\label{dndt}
\endeq

\subsection{Close-in companions}

In the same way, using the expression for the work given by eqn. (\ref{dotWstat}), we obtain for close-in companions in synchronous or stationary rotation,
\beq
<\dot{n}>\  \simeq\
 \frac{9n^2k_dMR^5}{2ma^5}\,(7 e^2 + S^2) \epp_2.
\label{dotnstat}\endeq
If we put $\epp_2=1/Q$ and $S=0$, we obtain the usual expression given by Peale and collaborators (see Sect. \ref{joint}).

If tidal lags are assumed to be proportional to the frequencies of the corresponding tidal wave, and if the synchronization assumption $\Omega=n$ is introduced in Eqn. (\ref{dndt}), we obtain, when $S=0$, the same result as given by Eqn. (3) of Mardling and Lin (2004) for the variation of the semi-major axis due to the tides in a synchronous companion \footnote[18] {{\it Note added in this version.} In fact, in order to have the agreement, we had to make $k_d\epp_2=k/2Q$ instead of $k_d\epp_2=k/Q$ as done to compare to other results of the same authors. If the latter factor is used also in this case, the coefficient in their Eqn. (3) should be $\frac{171}{2}$ instead of $\frac{171}{4}$.}. However, the result thus obtained does not take into account the impossibility of having $\Omega=n$ without the existence of an additional non-tidal torque counteracting the tidal torque. The consideration of this necessary torque reduces the term $19e^2$ appearing in the equations for $\dot{a}$ and $\dot{n}$ to $7e^2$.   

\section{Variation in Eccentricity and Inclination}\label{e-i} 

The orbital angular momentum of the system is given by ${\cal L}=|{\cal L}|\,\hat{\mathbf{u}}$ where ${\mathbf{u}}$ is the unit vector perpendicular to the orbital plane and 
\beq
|{\cal L}|=\frac{Mm}{m+M}na^2\sqrt{1-e^2}=\frac{GMm}{na}\sqrt{1-e^2}.
\endeq
$a$, $e$ are, respectively, the semi-major axis and eccentricity of the relative orbit (astrocentric).

The fact that the torque and the angular momentum are both perpendicular to the line of nodes allows the equation $\dot{\cal L}={\cal M}$ to be reduced to the plane defined by them. In that plane, the time derivative of the angular momentum is $\dot{\cal L}=\frac{d}{dt}|{\cal L}|\, \hat{\mathbf{u}} + |{\cal L}| \dot{I}\, \dot{\imath}\hat{\mathbf{u}}$ and, from the given equation, we obtain $\frac{d}{dt}|{\cal L}| = M_\perp$ and $|{\cal L}|\dot{I}= M_\parallel$, where $M_\perp$ and $M_\parallel$ are the components of ${\cal M}$ along the directions of $\hat{\mathbf{u}}$ and $\dot{\imath}\hat{\mathbf{u}}$.

\begin{figure}[t]
\centerline{\hbox{
\includegraphics[height=3.5cm,clip=]{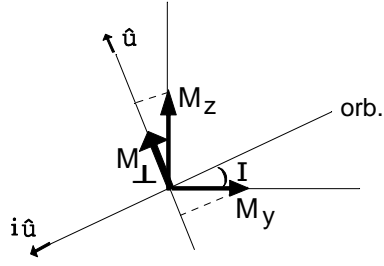}}}
\caption
{Meridian plane normal to the nodal line. Components of the torque along the $z$-axis  ($M_z$), projected on the reference plane ($M_{y}$) and perpendicular to the orbit $(M_\perp $). $\hat{\mathbf{u}}$ and $\dot{\imath}\hat{\mathbf{u}}$ are unit vectors. The orientation of $M_{y}$ 
in this figure corresponds to the case $\epp_0+\epp_8-\epp_9 < 0$ (see eqn. \ref{torques}). Note that $M_x=0$.}
\label{eplot4}    
\end{figure}

From figure \ref{eplot4}, we have
\begin{eqnarray}
M_\perp & \speq & \phantom{-} M_z \cos I - M_{y} \sin I \nonumber \\
M_\parallel &\speq & -M_z \sin I - M_{y} \cos I.
\end{eqnarray}

Solving the two components of the equation $\dot{\cal L}={\cal M}$, we obtain
\beq
<\dot{e}>\speq -\frac{3nek_d MR^5}{8ma^5}\big(
2\epp_0-\frac{49}{2}\epp_1+\frac{1}{2}\epp_2+3\epp_5\,\big)
\label{dedt}\endeq
and
\beq
<\dot{I}>\speq 
\frac{3nk_d SMR^5}{4ma^5}\big(-\epp_0+\epp_8-\epp_9\big)
\label{didt}\endeq

\subsection{Close-in companions}\label{J22prec}
In the case of close-in companions, we have $\epp_2=-\epp_1=-\epp_8=\epp_9>\ 0$. We also assume that $\epp_5=\epp_2$. (The lag $\epp_5$ corresponds to a radial tidal wave with same frequency as the monthly tide.) Hence
\beq
<\dot{e}> \speq  -\frac{3nek_d MR^5}{4ma^5}
\big(\epp_0+14\epp_2\,\big)
\endeq\beq
<\dot{I}> \speq  
-\frac{3nSk_d MR^5}{4ma^5}\big(\epp_0+2\epp_2\big)
\endeq

For synchronous or stationary companions we have $\epp_0\le {\cal O}(e^2)$ and, thus,
\beq
<\dot{e}> \speq  -\frac{21nek_dMR^5}{2ma^5} \epp_2
\label{dotestat}\endeq\beq
<\dot{I}> \speq  
-\frac{3nS k_d MR^5 }{2ma^5}\epp_2
\endeq

The results obtained are the same as found in the literature since Goldreich (1963),  Goldreich and Soter (1966) and Peale et al. (1980) if we consider the quality factor $Q$ as the inverse of the phase lag of the monthly tide (which coincides with the diurnal tide because of the synchronization)\footnote{For the conversion of Goldreich and Soter results (as well as for those of Dobbs-Dixon et al. (2004) and Mardling and Lin (2004)) we use $Q'=3Q/2k$.}. 

At the order considered in this paper, the only contribution of the torque
due to $U_{22}$ is a motion of the node:\footnote{Since $\omega$ is expected to grow monotonically due to the oblateness of the bodies, one part of the given motion of the node is not secular. We included it in the given equation for sake of completeness} 
\beq
{<\dot{\widehat{\rm{ON}}}_{22}>} \speq -\frac{3J_{22}nR^2}{a^2}\big(1-2e^2-\frac{1}{4}S^2-\frac{9}{4}e^2 \cos 2\omega\big).
\label{Node22}\endeq
However, this precession exists even when $\delta=0$. Therefore, it is not a tidal effect due to the misalignment of $U_{22}$, but one due to the figure of the body trapped into the 1:1 spin-orbit resonance.

\subsection{Variation in obliquity} 

The obliquity $I$ may vary due to the torque acting on the orbit and its counterpart acting on the deformed body. Thus, its variation is given by the sum of the two components: $<\dot{I}>$ and $<\dot{J}>$. From eqns. (\ref{dotJ}) and (\ref{didt}), we have
\beq
<\dot{I}+\dot{J}> \speq \frac{3k_dSGM^2R^5}{4C\Omega a^6}
\Big[\Big(1
-\frac{C\Omega an}{GMm}\Big)\epp_0 +
\Big(1+
\frac{C\Omega an}{GMm}\Big)\big(\epp_8-\epp_9\big)\Big].
\endeq
In the case of close-in companions, we have
\begdi
\frac{C\Omega an}{GMm} \simeq  \frac{C}{ma^2}\frac{\Omega}{n} \ll 1,
\enddi
showing that in this case only the part coming from $<\dot{J}>$ matters.
The contribution from $<\dot{I}>$ can be neglected.

\section {Tidal friction in the central body}\label{centralbody}

Darwin's treatment of tidal friction allowed us to obtain equations that are valid in all cases, provided that the tidal phase lags and response factors are left as free parameters. Equations (\ref{dndt}), (\ref{dedt}) and (\ref{didt}) may be used to describe the effects of tidal friction on the central body. However, we have to consider separately the case in which the central body is a fast-rotating body (as Jupiter or Saturn) or a slow-rotating body (as the Sun and many other main-sequence stars) .

In the forthcoming equations, one should remember that the parameters $R$ (radius), $m$ (mass), $k_d$ (dynamical Love number) refer to the deformed body, which is now the central body, while $M$ is the mass of the tide-raising body, the close-in companion. The phase lags $\ep_i$ also refer to the central body. 

\subsection{Type 1: $\Omega \gg n$. Fast-rotating planet}  

In this case, the terms appearing in the equations obtained thus far are semi-diurnal ($\epp_0$, $\epp_1$, $\epp_2$), diurnal ($\epp_8$, $\epp_9$) and monthly ($\epp_5$) (all positive). The equations for the variation of the elements in this case become

\beq
<\dot{n}> \speq - \frac{3n}{2a} <\dot{a}>\speq -\frac{9n^2k_dMR^5}{2ma^5}\Big[
\epp_0+e^2(\frac{27}{2}\epp_0-\frac{3}{4}\epp_5)-S^2(\epp_0-\epp_8)\Big]
\label{dotnty1}\endeq

\beq
<\dot{e}>\speq \frac{3nek_dMR^5}{8ma^5}\big(
22\epp_0-3\epp_5\,\big)
\label{dotety1}
\endeq
and
\beq
<\dot{I}>\speq 
-\frac{3nSk_dMR^5}{4ma^5}\,\epp_0 .
\endeq

The equations found usually in the literature (e.g. Yoder and Peale, 1981) are
\beq
<\dot{n}> \speq - \frac{3n}{2a} <\dot{a}>\speq
- \frac{9n^2k_dMR^5}{2mQa^5}\Big(1+ \frac{51}{4}e^2\Big)
\label{K-a}\endeq
\beq
<\dot{e}>\speq \frac{57 MR^5nk_d}{8mQa^5}\, e
\label{K-e}\endeq
which are obtained from the above equations after neglecting $S^2$, considering equal the lags and response factors of both diurnal and monthly (radial) tides and introducing $|\ep_0|=1/Q$. 
The coefficients in these expressions are slightly different from those found in Kaula(1964) and in Goldreich and Soter (1966) since those authors adopt here $M+m\simeq m$, which is equivalent to neglect the reaction force. There is also a sign difference in Kaula(1964), but it is due to the opposite definition of the corresponding 
phases.

\subsection{Type 3: $\Omega \ll n$. Slow-rotating star}  

In this case, the terms appearing in the equations obtained thus far are semi-annual ($\epp_0$, $\epp_8$), tierce-annual ($\epp_1$), annual ($\epp_2$, $\epp_5$) (all negative except $\epp_5$) and the ``diurnal" lag ($\epp_9$) (positive). 
The equations for the variation of the elements in this case become

\beq
<\dot{n}> \speq - \frac{3n}{2a} <\dot{a}> \speq -\frac{9n^2k_dMR^5}{2ma^5}\Big[
\epp_0-e^2(5\epp_0-\frac{147}{8}\epp_1-\frac{7}{8}\epp_2)\Big]
\label{dotnty3}
\endeq

\beq
<\dot{e}>\speq -\frac{3nek_dMR^5}{4ma^5}\big(
\epp_0-\frac{49}{4}\epp_1 -\frac{5}{4}\epp_2\,\big)
\label{dotety3}\endeq
where we have introduced $\epp_5=-\epp_2$ and $\epp_8=\epp_0$. We also have 
\beq
<\dot{I}>\speq 
-\frac{3nSk_dMR^5 }{4ma^5} \epp_9.
\endeq

The equivalent of equations (\ref{K-a}) and (\ref{K-e}) is obtained assuming that the annual, semi-annual and tierce-annual phase lags and Love numbers to be equal. They are 
 \beq
<\dot{n}> \speq - \frac{3n}{2a} <\dot{a}>\speq
\frac{9n^2k_dMR^5}{2mQa^5}
\Big(1+\frac{57}{4}e^2\Big)\label{dotAstar}
\endeq\beq
<\dot{e}>\speq -\frac{75MR^5nk_d}{8mQa^5}\,e \endeq
where we have introduced $|\ep_0|=1/Q$. 

One may note that, as expected from the opposite geometry presented by Types 1 and 3, the effects on the semi-major axis and eccentricity have different signs. However, they do not have the same magnitudes because some phase lags have different signs. The consideration of equal signs for all $\epp_i$ in Type 3 would be physically equivalent to put some waves with a phase advance with respect to the corresponding tidal waves in the static model (instead of a lag behind them).

\medskip

The above results are only valid if $\Omega \gg n$ or $\Omega \ll n$, respectively. For all intermediary cases, the general equations should be used.

\subsection{Rotation of the central body}

From eqns. (\ref{dotOmega}) and (\ref{dotJ}), introducing the lags in the same way as above, we obtain:

\subsubsection{Type 1 central body}

\beq
<\dot{\Omega}> \speq - \frac{3k_dGM^2R^{5}}{2Ca^6}\Big((1+ 
\frac{15}{2}e^2)\epp_0-S^2(\epp_0-\epp_8)\Big)
\label{dotOmegatipo1}\endeq

\beq
<\dot{J}>  \speq \frac{3k_dGM^2R^{5}}{4C\Omega a^6}S\epp_0 
\endeq

\subsubsection{Type 3 central body}

\beq
<\dot{\Omega}> \speq - \frac{3k_dGM^2R^{5}}{8Ca^6}\Big(4\epp_0 
+e^2(-20\epp_0+49\epp_1+\epp_2)+2S^2(-\epp_0+\epp_9)\Big)
\endeq
\beq
<\dot{J}> \speq \frac{3k_dSGM^2R^{5}}{4C\Omega a^6}(2\epp_0-\epp_9) 
\endeq

In the circular approximation, the results given by Goldreich and Soter (1966) for $<\dot{\Omega}>$ coincide with those given by eqns. (\ref{dotOmegaclose}) and (\ref{dotOmegatipo1}) if we adopt the same definitions for the moment of inertia used by Goldreich and Soter, i.e., $C=\alpha mR^2$ and $C=0.4\,MR^2$, respectively.

\section{Linear theories with a constant time lag}\label{linear}

Several of the existing tidal friction theories introduce \textit {ab initio} the tidal lag by assuming that the tides correspond to the position of the tide-raising body $\tens{M}$ at a time $\tau$ before the current time (see e.g. Mignard, 1979).  For the sake of comparing the results of linear theories to those obtained here, we give below the resulting equations when the $\epp_i$ are assumed to be proportional to the frequencies of the corresponding tidal waves, with $\tau$ as the coefficient of proportionality\footnote[19] {{\it Note added in this version.} This is equivalent to assume the lags proportional to the frequencies and the same dynamical response factors $k_d$ for all terms.}.

For the equations giving the variation of the rotational state of the deformed body, we obtain
\beq
<\dot{\Omega}> \speq  \frac{3nk_dGM^2R^{5}}{Ca^6}
\Big[\big(1+\frac{27}{2}e^2-\frac{1}{2}S^2\,\big)-\big(1+\frac{15}{2}e^2-\frac{1}{2}S^2\,\big)
\frac{\Omega}{n}\,\Big]\tau
\label{dotOmegaDobb}\endeq
\beq
<\dot{J}> \speq - \frac{3nk_dGM^2R^{5}}{C\Omega a^6}S(1-\frac{1}{2}\frac{\Omega}{n})\,\tau. 
\endeq
Eqn. (\ref{dotOmegaDobb}) reproduces eqn. (18) of Dobbs-Dixon et al. (2004) to the adopted order in eccentricity, except for modifications done there using  Kepler's third law, the effects due to stellar wind and some specific considerations on the parts of the body participating in the angular momentum exchange . 
The so-called quality factor used there is related to the parameters used here through $Q=1/n\tau$.

For the equations giving the variation of the orbital elements, we obtain

\beq
<\dot{n}> \speq -\frac{3n}{2a}<\dot{a}> \speq \frac{9n^3k_dMR^5}{ma^5}
\Big[\big(1+ 23 e^2\, \big) -
\big(1+\frac{27}{2}e^2-\frac{1}{2}S^2\, \big)\frac{\Omega}{n}\Big]\tau,
\label{dotnDarwin}\endeq

\beq
<\dot{e}>\speq -\frac{27n^2e k_d MR^5}{ma^5}
\big(1-\frac{11}{18}\frac{\Omega}{n}  \big) \tau,
\label{dedtDobb}\endeq

\beq
<\dot{I}>\speq -
\frac{3nk_d SMR^5}{2ma^5}\Omega\tau.
\label{didtOmega}\endeq

These equations are the same found in several papers on tidal friction on close-in exoplanets using Hut's approach (e.g. Mardling and Lin, 2004, Dobbs-Dixon et al. 2004). It is worth noting that the results in Hut(1981) are not given by expansions but by closed formulas.

\section{Cumulative orbital variations due to tides in both bodies}\label{joint}%

In this section, we add the variations of the mean-motions and eccentricities due to the tides raised in both the central body and the companion.
In the equations giving the variations due to the tides in the central body, we make the substitutions $k_d=k_{dA}$, $\epp_j=\epp_{jA}$, $m=m_A$, $M=m_B$, $R=R_A$, $S=S_A$ and $\Omega=\Omega_A$. In the equations giving the variations due to the tides in the synchronous or stationary companion, we make the substitutions $k_d=k_{dB}$, $\epp_j=\epp_{jB}$, $m=m_B$, $M=m_A$, $R=R_B$, $S=S_B$ and $\Omega=\Omega_B$.

For sake of simplicity, we introduce the factor 
\beq
D\speq \frac{k_{dB}}{k_{dA}}\left|\frac{\epp_{2B}}{\epp_{0A}}\right|\Big(\frac{m_A}{m_B}\Big)^2\Big(\frac{R_B}{R_A}\Big)^5.
\endeq
This is the same factor introduced in Yoder and Peale (1981) if we assume that $k_{dA}$ and $k_{dB}$ are the Love numbers associated to the tidal waves whose frequencies are $2\Omega_A-2n$ and $2\Omega_B-n$ and introduce the ratio of the phase lags instead of the ratio of the quality factors. The conversion is done using $|{\ep_{2B}}/{\ep_{0A}}| = {Q_A}/{Q_B}$. 

Let us initially add the variations of the mean-motions and eccentricities given by eqns. (\ref{dotnty1}), (\ref{dotnstat}), (\ref{dotety1}) and (\ref{dotestat}) corresponding to a {\underline{fast-rotating central planet}} and a satellite in stationary rotation.

If we assume that the lags are independent of the frequencies and equal, we obtain
\beq
<\dot{n}> \speq 
-\frac{9n^2k_{dA}m_BR_A^5\epp_{0A}}{2m_A a^5}
\Big(1+\frac{51}{4}e^2 - D(7 e^2 + S_B^2)\Big)
\label{dotnjoint}\endeq
\beq
<\dot{e}>\speq \frac{3nek_{dA}m_BR_A^5\epp_{0A}}{2m_Aa^5} \big(\frac{19}{4}-7 D\big)
\label{dotejoint}\endeq

Inserting $k_{dA}\epp_{0A}=k_A/Q_A$ and $S_B=0$ into eqns. (\ref{dotnjoint}) and (\ref{dotejoint}), they become the ones often appearing in the literature on satellites' tidal friction and are the same used by Peale and co-workers (e.g. Yoder and Peale, 1981).

In the case of linear theories with a constant time lag, phase lags are proportional to the corresponding tidal wave frequencies and the equations for the variation of the elements become
\beq
<\dot{n}> \speq 
-\frac{9n^2k_{dA}m_BR_A^5\epp_{0A}}{2m_A a^5}
\Big(1+\frac{54}{4}e^2 -\frac{1}{2}S_A^2 - D(7 e^2 + S_B^2)\Big)
\endeq
\beq
<\dot{e}>\speq \frac{3nek_{dA}m_BR_A^5\epp_{0A}}{2m_Aa^5} \big(\frac{11}{2}-7 D\big).
\endeq
It is worth noting that there is some robustness in the last two sets of equations due to the fact that for fast-rotating central bodies $\epp_0=\epp_1=\epp_2$. The only difference between the two sets comes from the contribution of the monthly (radial) tide ($\epp_5$) in the central body and the non-vanishing term in $S_A^2$, which do not appear in the first set of equations because of the assumption of the same phase lag for the diurnal and semi-diurnal tidal waves.   
Let us now do the same for a {\underline{slow rotating star}} and an exoplanet in stationary rotation, but given the diversity of frequencies in this case, we consider only the case of tidal lags proportional to the corresponding tide frequencies. We expand eqns. (\ref{dotnDarwin}) and (\ref{dedtDobb}), to the order ${\cal O}(\Omega_A/n)$ $(\ll 1)$ and assume $\Omega_B\simeq n(1+6e^2)$. Hence (correcting the typos of the previous version):

\beq
<\dot{n}> \speq 
\frac{9n^2k_{dA}m_BR_A^5|\epp_{0A}|}{2m_A a^5}
\Big(1+23e^2+D(7 e^2 + S_B^2) +\frac{\Omega_A}{2n} 
(19e^2+S_A^2)\Big) 
\endeq
\beq
<\dot{e}>\speq -\frac{27nek_{dA}m_BR_A^5|\epp_{0A}|}{2m_Aa^5}
\Big(1+ \frac{7}{9} D + \frac{7}{18}\frac{\Omega_A}{n}\Big)
\endeq

In all cases, if the satellite or exoplanet is in synchronous instead of stationary rotation, we have the same equations. However, the giant planets of the Solar System do not show a measurable $J_{22}$ and we should not expect a permanent equatorial ellipticity in hot Jupiters. Therefore, hot Jupiters with quality factor not too large will have a fast synchronization but will not reach exact synchronization before circularization. They will be rather driven to a stationary rotational state. Anyway, in both cases the equations giving tidal variations of the orbital elements are the same.
The term in $19De^2$ sometimes found in equations for the tidal variation of the semi-major axis and mean-motion of close-in exoplanets does not take into account that a synchronous rotation is only possible if an external torque counteracts the tidal torque and may be substituted by $7De^2$ as given above. The only difference is that, in the case of synchronous rotations, we will have to consider also the non-tidal effects due to the equatorial asymmetry of the companion.

\section{Motions with respect to an invariable plane}

Throughout the previous sections, we have considered the relative positions of the equatorial and orbital planes. None of these planes is fixed. In more general situations (as considered in the next section), it may become necessary to refer their positions to an inertial frame. In that case, we may adopt the invariable plane (normal to the total angular momentum) as reference plane. If all poles are represented on the celestial sphere, the equatorial and orbital poles may be considered as three points whose ``masses" are the corresponding angular momenta and whose gravity center lies on the pole (projected in the origin) of the invariable plane. The distances of the poles of the two equators to the pole of the orbit are the mutual inclinations (obliquities) $I$. The angle between the arcs joining the pole of the orbit to the pole of the two equators is the distance of the nodes of the two equators on the orbital plane. These data are sufficient to completely determine all parameters of the considered planes. 

\begin{figure}[h]
\centerline{\hbox{
\includegraphics[height=3.5cm,clip=]{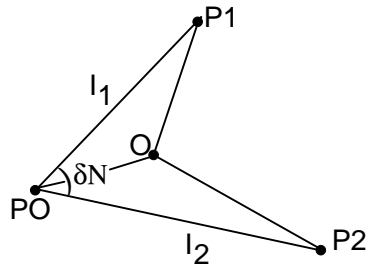}}}
\caption
{Projection of the orbital (${\rm{P_O}}$) and equatorial ($\rm{P_1, P_2}$) poles on the invariable plane. $\delta N$ is the angle between the nodal lines.
$I_1$ and $I_2$ are the two obliquities. }
\label{eplot6}    
\end{figure}

We have to consider two steps. The first one is to determine the positions of the orbital (${\rm{P_O}}$) and equatorial ($\rm{P_1, P_2}$) poles with respect to the pole (O) of the invariable plane when we know $\delta N, I_1,I_2$. If $M_0, M_1, M_2$ are the angular momenta, respectively, orbital and equatorial, and if we denote by ${\cal J}_0,{\cal J}_1,{\cal J}_2$ the distances (angles) of the corresponding poles to the pole of the invariable plane, we have:
\begin{eqnarray}
&M_1\sin \widehat{\rm P_1OP_O} - M_2 \sin \widehat{\rm P_2OP_O} = 0 &\\ \nonumber
&M_1\cos \widehat{\rm P_1OP_O} + M_2 \cos \widehat{\rm P_2OP_O} + M_0 = 0 & 
\end{eqnarray}
\begin{eqnarray}\label{triang}
\cos I_1 &\speq& \cos {\cal J}_0 \cos {\cal J}_1+\sin{\cal J}_0\sin{\cal J}_1\cos \widehat{\rm P_1OP_O}
\\ \nonumber
\cos I_2 &\speq& \cos {\cal J}_0 \cos {\cal J}_2+\sin{\cal J}_0\sin{\cal J}_2\cos \widehat{\rm P_2OP_O}\end{eqnarray} 
and 
\beq
\cos I_1 \cos I_2 + \sin I_1 \sin I_2 \cos \delta N = 
\cos {\cal J}_1 \cos {\cal J}_2 + \sin {\cal J}_1 \sin {\cal J}_2 \cos \widehat{\rm P_1OP_2}\label{bitriang} 
\endeq
which are enough to determine the arcs ${\cal J}_0,{\cal J}_1,{\cal J}_2$ and the angles between them. 

The converse step is to determine $I_1,I_2, \delta N$ when the position of the poles with respect to the pole of the invariable plane are known. This can be done easily using eqns. (\ref{triang} -- \ref{bitriang}).

The possible polar oblateness of the deformed body was not considered in previous sections. Indeed, a torque component directed along the nodal line does not change the given results. However, it will make the nodal line precess and such precession (a uniform variation of $\delta N$) cannot be neglected when considering eqn. (\ref{bitriang}).

\section{Equations of Motion}

We reproduce below the classical form of the equations of motion of two planets, including the forces due to tidal friction.

Let us consider two bodies with masses $m_A$ (primary) and $m_B$ (companion).

Let $\mathbf{f}_0$ be the attractive force acting on $m_B$ due to $m_A$ in a point mass model and $-\mathbf{f}_0$ its reaction acting on $m_A$;

Let $\mathbf{f}_A$ be the tidal force acting on $m_A$ due to the tidal deformation of $m_B$. It is given by eqns. (\ref{f1ueva} -- \ref{f3ueva}) making the substitutions $k_d=k_{dB}$, $M=m_A$, $R=R_B$, $S=S_B$ (sine of the obliquity of the equator of $m_B$), $\epp_j=\epp_{jB}$ and $\omega=\omega_B$ (argument of the periapsis reckoned from the intersection of the orbit and the equatorial plane of B). Let $-\mathbf{f}_A$ be the corresponding reaction acting on $m_B$.

Let $\mathbf{f}_B$ be the tidal force acting on $m_B$ due to the tidal deformation of $m_A$. It is given by eqns. (\ref{f1ueva} -- \ref{f3ueva}) making the substitutions $k_d=k_{dA}$, $M=m_B$, $R=R_A$, $S=S_A$ (sine of the obliquity -- inclination over the orbital plane -- of the equator of $m_A$), $\epp_j=\epp_{jA}$, $\omega=\omega_A$ (argument of the periapsis reckoned from the intersection of the orbit and the equatorial plane of A). Let $-\mathbf{f}_B$ be the corresponding reaction acting on $m_A$.

The equations of the motion with respect to an inertial reference frame are:
\begin{eqnarray}
m_A \ddot\mathbf{r}_A &\speq & -\mathbf{f}_0 + \mathbf{f}_A - \mathbf{f}_B\\ \nonumber
m_B \ddot\mathbf{r}_B &\speq & \phantom{-}\mathbf{f}_0 - \mathbf{f}_A + \mathbf{f}_B.
\end{eqnarray}

Let, now, $\mathbf{r} = \mathbf{r}_B -\mathbf{r}_A$ be the radius vector of the companion in a reference frame fixed in the primary. From the above equations we obtain
 
\beq
m_B \ddot\mathbf{r} \speq (1+\frac{m_B}{m_A})\big(\mathbf{f}_0 - \mathbf{f}_A + \mathbf{f}_B\big). \label{eqmov}
\endeq

The only critical step comes from the fact that the forces $\mathbf{f}_A$ and $\mathbf{f}_B$ were given in different reference systems. In previous sections, all results corresponded to scalar quantities -- work, semi-major axis, eccentricity, obliquities -- which are not affected by the frames being used, but, in the above equation, it is necessary to take this into account. If $[\mathbf{f}_A]$ and $[\mathbf{f}_B]$ represent the vectors whose components are given by eqns. (\ref{f1ueva} -- \ref{f3ueva}), we have to change eqn. (\ref{eqmov}) into 

\beq
m_B [\ddot\mathbf{r}] \speq (1+\frac{m_B}{m_A})\big([\mathbf{f}_0] - e^{i\pi}[\mathbf{f}_A] + [\mathbf{f}_B]\big).
\endeq
where $e^{i\pi}$ indicates a $\pi$-rotation around the axis normal to the orbital plane. For all other vectors, $[..]$ represent their components in the system with center in $m_A$. Note that because of the adopted spherical coordinates frames, we have $e^{i\pi}[a,b,c]^T=[-a,b,-c]^T$.
 
The other point to be taken into consideration is the relationship between the angle variables in both cases. The mean anomaly $\ell$ does not depend on the adopted frame, but the arguments of the periapsis: $\omega_k$ are different and their difference is the angle $\delta N$ introduced in the previous section.

\subsection{Disturbing Function in perturbation equations}\label{corrGauss}

Many theories since Darwin's (e.g. Kaula, 1964, MacDonald, 1964) use the  classical Gauss or Lagrange equations for the variation of the orbital elements. In order to assess the consequences here of eqn. (\ref{eqmov}), let us consider, for simplicity, the particular case in which the companion of mass $m_B$ is a mass point that cannot be deformed. 
In such case, the equations of motion are simply reduced to $m_B \ddot\mathbf{r} \speq (1+\frac{m_B}{m_A})(\mathbf{f}_0 + \mathbf{f}_B)$. The central force $\mathbf{f}_0$ derives from the two-body potential $U_0$
and the perturbation $\mathbf{f}_B$ derives from a potential ${\cal R}$, i.e. $\mathbf{f}_B=-m_B \, {\rm grad} \,{\cal R}$. The disturbing potential ${\cal R}$ appearing in Gauss and Lagrange equations corresponds to an  ``external" perturbation acting on $m_B$. However, in the present case, the perturbation acting on $m_B$ is ``internal" to the system of bodies. Therefore, as is  usually done in the formulation of an N-body problem, the reactions must also be taken into account, that is, ${\cal R}$ in Gauss and Lagrange equations has to be substituted by $(1+\frac{m_B}{m_A}){\cal R}$ to take into account the reaction on $m_A$ of its tidal action on $m_B$. This correction can be neglected only when $m_B \ll m_A$, which is the case in the cited theories. 

For sake of completeness, we recall that the reaction to $\mathbf{f}_0$ is duly taken into account in the current formulation of Gauss and Lagrange equations. 

\section{Conclusions}

We have done a review of Darwin's theory for bodily tides and compared its results to those published in many papers dealing with tidal evolution of orbital elements and obliquities and the energy dissipation inside bodies due to tidal friction. As expected, our main results coincide with the general results given by Kaula (1964) to the order of the approximations adopted. The only differences are that we present the results through explicit formulas and that we have not introduced approximations as $m_A+m_B \simeq m_A$; the given equations remain valid even in the case of larger mass ratios as in the Earth-Moon system and in the possible case of a close binary system formed by a normal star and a big brown dwarf. 

The comparison to more recent results allowed us to know, in each case, the additional hypotheses made by the authors. In some cases the formulas used (e.g. Yoder and Peale, 1981) correspond to assuming tidal lags independent of frequencies (as done in eqns. \ref{dotnjoint} and \ref{dotejoint}) while others (e.g. Mardling and Lin, 2004) correspond to adopting the same linear relationship between lags and frequencies introduced by Darwin (as done in Sec. \ref{linear}). 
Other consequences are the differences between the equations for the variation of the orbital elements in the cases planet-satellite and star-planet, due to the different rotational state of the central body in these problems. 

The agreement between the various theories is limited to terms corresponding to the circular approximation. Many disagreements are found in terms depending on the eccentricity\footnote{Other disagreements are pointed out by Efroimsky (2008).}
showing that it is necessary to improve our knowledge of the physics of the processes leading to tidal lags before extending the results to higher orders in eccentricity and inclination. In practical applications, these divergences are absorbed in the poorly known quality factor $Q$ and results of evolution studies do not diverge in the same proportion.

Emphasis is given in the paper to the case of companions having reached one of the two possible final states of the companion synchronization. If tides are the only source of perturbations in the system, perfect synchronization cannot be achieved while the orbit is not completely circularized. If a remnant eccentricity exists, the final state is rather non-synchronous, the rotation reaching a stationary state with a rotation velocity slightly larger than the orbital mean-motion. In order to have true synchronous rotation, it is necessary to provide the system with a torque counteracting the tidal torque. This additional torque can be provided by a permanent solid-like equatorial asymmetry of the companion. Tides drive the figure of the body to a small misalignment and the misalignment provides one torque acting in opposition to the tidal torque. The results for the tidal variations on the system are the same in the stationary and synchronous cases, but, in the synchronous case, we have several non-tidal effects coming from the figure of the body which may change the evolution of the system. Some equations are found in the literature using the condition $\Omega=n$, which corresponds to the truly synchronous case, but without taking into account the counteracting torques allowing for  the synchronization. Those equations are incomplete. 

Darwin's harmonic decomposition of the tide allowed us to identify the role played in dissipation by different tidal waves. In general, dissipation is dominated by the semi-diurnal tide: the harmonic whose period is half the synodic rotation period. However when the rotation is captured in the 1:1 spin-orbit resonance and the synodic rotation period becomes infinite, dissipation is mainly due to the monthly tide: the harmonic whose period is the orbital period, which is now equal to the sidereal rotation period. In this case the dissipation becomes proportional to the square of the eccentricity.

\begin{acknowledgements}

The authors thank Drs. C.Beaug\'e, B. Jackson, J.Laskar, T.A.Michtchenko and J. Wisdom for their suggestions and comments. The authors acknowledge the support of this project by CAPES, CNPq and FAPESP (Brazil), the DFG (Deutsche Forschungsgemeinschaft) and the CAPES/SECYT joint program. We thank the three anonymous referees for their stimulating reports and the co-editor of this issue for handling this paper.

\end{acknowledgements}

\section*{Appendices}

\appendix
\section{Equatorial prolateness of Roche ellipsoids}

Roche ellipsoids are figures of equilibrium of a homogeneous fluid  with rotation synchronous to their orbital (circular) motion about a spherical companion. The analysis of the equilibrium equations shows, however, that the angular orbital velocity of the system and the angular rotation velocity of the body appear separately in the equations. Therefore the equilibrium equations remain valid even if there is no spin-orbit synchronization. 
The resulting equations in this more general case are (Jeans, 1929, sec. 206):
\begin{eqnarray} 
2\pi G\rho a^3 b c \int_0^\infty \frac{dx}{\Delta(a^2+x)} -\frac{2GMa^2}{r^3}-a^2\Omega^2 &\speq & \Theta\\ \nonumber
2\pi G\rho a b^3 c \int_0^\infty \frac{dx}{\Delta(b^2+x)} +\frac{ GMb^2}{r^3}-b^2\Omega^2 &\speq & \Theta\\ \nonumber
2\pi G\rho a b c^3 \int_0^\infty \frac{dx}{\Delta(c^2+x)} 
+ \frac{GMc^2}{r^3} &\speq & \Theta \nonumber
\end{eqnarray}
(except for terms of the order of the eccentricity)
where $\rho$ is the density, $r$ is the distance between the two centers, $\Omega$ is the angular rotation of the body and $M$ is the mass of the tide generating body.
The axes of the ellipsoid are $a$, $b$, $c\,$ ($c<b<a$) and $\Delta^2={(a^2+x)(b^2+x)(c^2+x)}$. 
$\Theta$ is an auxiliary function, the definition of which is not explicitly required in this context.

The shape of the ellipsoid equator is obtained by eliminating $\Theta$ between the two first equations: 
\beq
2\pi G\rho a b c \int_0^\infty \frac{(a^2-b^2)xdx}{\Delta(a^2+x)(b^2+x)}   \speq \frac{GM(2a^2+b^2)}{r^3} + (a^2-b^2)\Omega 
\endeq
If the oblateness of the ellipsoid is small, the integrals may be evaluated by means of some trivial expansions (Tisserand, 1891) to give, to the first order of approximation in the equatorial prolateness:
\beq
2\pi G\rho a b c \int_0^\infty \frac{(a^2-b^2)xdx}{\Delta(a^2+x)(b^2+x)}   \speq \frac{4Gm}{5R}\,\epsilon.
\endeq 
($R$ is the mean radius and $m$ is the mass of the ellipsoid). At the same order of approximation, we have $(a^2-b^2)=2R^2\epsilon$. We also adopt the approximation $2a^2+b^2=3R^2$ (the corrections of the order of the polar oblateness may be discarded because of the multiplication of this term by the small quantity $1/r^3$). Hence
\beq
\epsilon =\frac{15}{4}\Big(\frac{M}{m}\Big)\Big(\frac{R}{r}\Big)^3\Big
(1-\frac{5(M+m)}{2m}\frac{\Omega^2}{n^2}\frac{R^3}{r^3}\Big)^{-1}
\endeq

\section{Variations in the argument of periapsis and node\protect\footnote{With the corrections published in the Errata. The results given in the paper are incomplete. To obtain complete second-degree results in $e, I$, it is necessary to consider in $U_2$ terms up to fourth degree. In particular, it is necessary to use 
$$
\varphi\simeq v+\omega-\frac{1}{4}\sin(2v+2\omega)\sin^2 I
-\Big(\frac{1}{8}\sin(2\omega+2v)-\frac{1}{32}\sin(4\omega+4v)\Big) \sin^4 I.
$$
instead of the approximation given in eqn. (8), as well as higher-order equations for the Keplerian motion.\protect\\ 
\protect\phantom{m} The higher-order terms thus introduced also contribute the variations in the node given in the added sub-section. } }

For sake of completeness, we should also consider the variations in the argument of periapsis and the time of periapsis. They can be easily obtained from the corresponding Gauss equations
(see Beutler, 2001, p.230).
The force components $F_{\rm R}$, $F_{\rm S}$, $F_{\rm W}$ (radial, transverse and normal)\footnote{Usually these equations are written in terms of the accelerations $R,S,W$ instead of the forces} appearing in Gauss equations are related to the forces $F_1$, $F_2$, $F_3$ through the same rotation shown in fig. \ref{eplot3}:\footnote{The second- and third-degree contributions to $F_1,F_2, F_3$ are given in section \ref{forces} and Appendix D of the astro-ph version of the paper.} 
\begin{eqnarray}
F_R\,&=&\phantom{-}F_1\nonumber\\
F_S\;&=&-F_2\sin\beta+F_3\cos\beta\\
F_W&=&-F_2\cos\beta-F_3\sin\beta\nonumber
\end{eqnarray}
We introduce these relations in the corresponding Gauss equation multiplied by $(1+M/m)$ as discussed in section \ref{corrGauss} and average the result over an orbital period. We note that, at variance with the other perturbations, in the case of the argument of perihelion we have to consider both the dynamic and the static tide. Hence
\begin{eqnarray}
<\dot{\omega}_{\rm lag}> &\speq & \frac{9k_dnMR^5}{16ma^5}
\Big{[}(3\epp_0-\epp_2-\epp_5+3\epp_6-3\epp_8-\epp_{15}+\epp_{16}+3\epp_{25})\,e^2 \nonumber \\ 
&& \phantom{mmmmmm} - \frac{1}{2}
(6\epp_0-\epp_2-\epp_5-6\epp_7-12\epp_8+\epp_{11}+\epp_{12}-2\epp_{15}+2\epp_{16})\,S^2
\Big{]}\sin2\omega \nonumber 
\end{eqnarray}
and
\beq
<\dot{\omega}^U> \speq \frac{15k_fnMR^5}{2ma^5}\left(1+\frac{13}{2}e^2 \right),
\endeq
respectively. The comparison of them shows that the part coming from the dynamic tide is attenuated by the product by the lags $\epp_i$ and by $e^2$ and $S^2$ so that the motion of the periapsis is dominated by the static tide. The new lags appearing in the above expressions are given below:

\smallskip

\begin{center}\begin{tabular}{c|c}
  Subscript No. & Frequency  \\
  \hline\\
  11 & $2\Omega-n $  \\
  12 & $n $  \\
  14 & $\Omega+n$  \\
  15, 16 & $\Omega-n$  \\
  17 & $\Omega -3n $ \\ 
  25 & $\Omega-2n$
\end{tabular}\end{center}
\smallskip

The quotient of the contribution from the tide raised in a close-in companion to that coming from the tide raised in the central body is 
\begdi
\frac{k_{dB}}{k_{dA}}\big(\frac{\overline{\rho}_A}{\overline{\rho}_B}\big)^2 \frac{R_A}{R_B}
\enddi
where $A, B$ represent the central body and the companion respectively and $\overline{\rho}$ are the mean densities.
It shows that the effects of the tides raised on the companion dominate $<\dot{\omega}>$. 

The comparison of the tidal motion of the periapsis and the mean-motion is roughly given by $<\dot{\omega}^U_B/n>$. It is of the order ${\cal O}(R_B^2/a^2)$. (For this estimate we assumed similar densities and $R_A\sim a$.)
 
In the case of one companion in synchronous rotation, we have to consider also the contribution coming from the asymmetry of the companion. If we proceed in the same way as above, but using $U_{22}$ instead of $U_2$, we obtain
\beq
<\dot{\omega}_{22}> \speq -\frac{3nR_B^2J_{22}}{4a^2}\,
\big(16-15e^2-7S^2+9(e^2-S^2)\cos 2\omega\big)
\endeq
whose main part is again of the order ${\cal O}(R_B^2/a^2)$. However, this result is given only to show the order of magnitude of the results. The main contribution from the shape of the companion is expected to come from the zonal term proportional to $J_2 P_2(\cos\theta)/r^3$, which has not been included in the calculations presented in this report.

\subsection{Variations in the Node} 

At the same approximation, the motion of the node is given by
\beq
<\dot{\widehat{\rm {ON}}}_{\rm lag}> \speq -\frac{9k_dnMR^5}{16ma^5}
(3\epp_0-\epp_2-\epp_5+3\epp_6-3\epp_8-\epp_{15}+\epp_{16}+3\epp_{25})\,e^2 \sin2\omega.
\endeq

Because of the rotational symmetry around the axis $a$ (see Fig. \ref{eplot1}), the static tide does not contribute a variation of the node.

\section{Forces arising from $U_{\rm lag}$\protect\footnote{The appendices C, D, E were published only in the astro-ph version of the paper}}

\begin{eqnarray}\label{f1ulag}
F_{1}^{\rm lag}&=&-\frac{9k_d G M M^\ast R^{5}}{8a^3r^{\ast 4}}\Big{[}
P^2  \epp_0  \Big(2 - 5 e^2  - S^2\Big)\sin(2 \varphi^{\ast}-2\ell-2\omega)\nonumber\\&&
+ e P^2    (1 - \frac{1}{2} S^2) \Big(7\epp_1\sin(2 \varphi^{\ast} -3 \ell -2\omega)
- \epp_2 \sin(2 \varphi^{\ast}- \ell -2\omega)\Big)\nonumber\\&&
+{17} e^2  P^2  \epp_3 \sin(2 \varphi^{\ast} -4 \ell -2\omega)
+  P^2 S^2  \epp_4 \sin 2 \varphi^{\ast} \nonumber\\&&
- e \epp_5  (4-6 P^2 + 3 P^2 S^2)\sin \ell 
-{3} e^2  \epp_6 (2-3 P^2 )\sin 2\ell \nonumber\\&& 
+  P^2 S^2 \epp_7 \sin(2\ell + 2\omega)
+2QS \Big(\epp_8 \cos( \varphi^{\ast} - 2 \ell -2\omega) 
- \epp_9 \cos \varphi^{\ast} \Big)\nonumber\\&& 
+ \frac{1}{2}e  P^2 S^2\Big(  3  \epp_{10} \sin( 2 \varphi^{\ast}+\ell )
+  3  \epp_{11} \sin( 2 \varphi^{\ast}-\ell )\nonumber\\&&
\hskip 2cm -  \epp_{12} \sin(\ell + 2\omega) 
+ 7\epp_{13}  \sin(3\ell + 2\omega)\Big)\nonumber\\&&
- Q e S \Big( 3\epp_{14} \cos( \varphi^{\ast}+\ell) 
+  3\epp_{15} \cos(\varphi^{\ast} -\ell )\nonumber\\&&
\hskip 2cm+  \epp_{16} \cos(\varphi^{\ast} - \ell -2\omega) 
- 7\epp_{17} \cos(\varphi^{\ast} -3 \ell -2\omega) \Big) \Big]
\end{eqnarray}

\begin{eqnarray}\label{f2ulag}
F_{2}^{\rm lag}&=&\frac{3k_dGMM^{\ast} R^{5}}{8a^3r^{\ast 4}}\Big{[}
Q  \epp_0  \Big(2 - 5 e^2 \Big)\sin(2 \varphi^{\ast}-2 \ell -2\omega)\nonumber\\&&
+\frac{1}{2} e Q \Big(7\epp_1\sin(2 \varphi^{\ast} -3 \ell -2\omega )
- \epp_2 \sin(2 \varphi^{\ast}- \ell - 2\omega)\Big)\nonumber\\&&
+{17} e^2  Q  \epp_3 \sin(2 \varphi^{\ast} -4 \ell -2\omega)
+ 6eQ \epp_5  \sin \ell
+9 e^2 Q \epp_6 \sin 2\ell
\nonumber\\&&
+ 4(1-2P^2)S \Big(\epp_8 \cos( \varphi^{\ast} - 2 \ell -2\omega) 
- \epp_9 \cos \varphi^{\ast}\Big)\nonumber\\&& 
- 2(1-2P^2) e S \Big( 3\epp_{14} \cos( \varphi^{\ast}+\ell ) 
+  3\epp_{15} \cos(\varphi^{\ast} -\ell )\nonumber\\&&
\hskip 2cm+  \epp_{16} \cos(\varphi^{\ast} - \ell -2\omega) 
- 7\epp_{17} \cos(\varphi^{\ast} -3 \ell -2\omega ) \Big) \Big]
\end{eqnarray}

\begin{eqnarray}\label{f3ulag}
F_{3}^{\rm lag}&=&\frac{3k_dGMM^\ast R^{5}}{8a^3r^{\ast 4}}\Big{[}
2P  \epp_0  \Big(2 - 5 e^2  - S^2\Big)\cos(2 \varphi^{\ast}-2\ell -2\omega)\nonumber\\&&
+ 2e P    (1 - \frac{1}{2} S^2) \Big(7\epp_1\cos(2 \varphi^{\ast} -3 \ell -2\omega)
- \epp_2 \cos(2 \varphi^{\ast}- \ell -2\omega )\Big)\nonumber\\&&
+{34} e^2  P  \epp_3 \cos(2 \varphi^{\ast} -4 \ell -2\omega)
+  2P S^2  \epp_4 \cos 2 \varphi^{\ast}\nonumber\\&&
-2\frac{Q}{P}S \Big(\epp_8 \sin( \varphi^{\ast} - 2 \ell -2\omega) 
- \epp_9 \sin \varphi^{\ast}\Big)\nonumber\\&& 
+ e  P S^2\Big(  3  \epp_{10} \cos( 2 \varphi^{\ast}+\ell)
+  3  \epp_{11} \cos( 2 \varphi^{\ast}-\ell )\Big)\nonumber\\&&
+ \frac{Q}{P} e S \Big( 3\epp_{14} \sin( \varphi^{\ast}+\ell  ) 
+  3\epp_{15} \sin(\varphi^{\ast} -\ell )\nonumber\\&&
\hskip 2cm+  \epp_{16} \sin(\varphi^{\ast} - \ell -2\omega) 
- 7\epp_{17} \sin(\varphi^{\ast} -3 \ell -2\omega ) \Big) \Big]
\end{eqnarray}
We have used here $\frac{d Q}{d \theta^{\ast}} = 2-4 P^2$ and $\frac{d P^2}{d\theta^{\ast}} = Q$. The latest derivative leads to the formation of several terms proportional to $QS^2$, which were dropped out from the expression of  $F_{2}^{\rm lag}$ (but included in the third-degree expressions of Appendix D).

\section{Third-degree forces}

The terms of third and fourth degree of $U_2$ are:

\begin{eqnarray}
\delta_3 U_2&=&-\frac{3 k_f G M R^5}{8 a^3 r^{\ast 3}}\Big{[}
\frac{1}{8} e^3 \Big(18\cos\ell +\frac{106}{3}\cos 3\ell + \cos(2 \varphi^{\ast}- \ell-2\omega)-123 \cos(2 \varphi^{\ast}-3\ell-2\omega) \nonumber\\
&&
\phantom{-\frac{3 k_f G M R^5}{8 a^3 r^{\ast 3}}\frac{1}{8} e^3}
+\frac{1}{3}\cos(2 \varphi^{\ast}+ \ell-2\omega) 
+\frac{845}{3} \cos(2 \varphi^{\ast}-5  \ell -2\omega )\Big)\nonumber\\
&&
+\frac{1}{2} eS^2 \Big(
- 7\cos(2\varphi^{\ast}-3\ell-2  \omega)
+ \cos(2 \varphi^{\ast}- \ell-2\omega)
- 6 \cos\ell \\
&&
\phantom{+\frac{1}{2} eS^2} 
+3\cos(2 \varphi^{\ast}+ \ell)+
3 \cos(2 \varphi^{\ast}-\ell )
-\cos( \ell+2\omega) +7 \cos(3  \ell +2\omega )\Big)\nonumber\\
&&
+ eQS \Big(\sin(\varphi^{\ast}- \ell-2\omega)
+ 3  \sin(\varphi^{\ast}+ \ell ) \nonumber
+ 3\sin(\varphi^{\ast}- \ell )
- 7 \sin(\varphi^{\ast}-3  \ell - 2\omega )\Big)\Big{]}
\end{eqnarray} 
and
\begin{eqnarray}
\delta_4 U_2&=&-\frac{3 k_f G M R^5}{8 a^3 r^{\ast 3}}\Big{[}
\frac{1}{12}e^4\big(15 + 28\cos 2\ell +77\cos 4\ell\big) 
+\frac{1}{8}e^4 \big(13\cos(2\varphi^{\ast}-2\ell-2\omega)
 \\
&&\phantom{mmmmmm}
-\frac{920}{3}\cos(2\varphi^{\ast}-4\ell-2\omega) \nonumber
+533\cos(2\varphi^{\ast}-6\ell-2\omega)
+\frac{2}{3}\cos(2\varphi^{\ast}+2\ell-2\omega)\big) \\
&&
+\frac{1}{4}e^2 S^2\big(-6
-18\cos 2\ell\nonumber
-10\cos(2\ell+2\omega)
+34\cos(4\ell+2\omega)
+6\cos 2\varphi^{\ast}\\
&&\phantom{mmmmmm}
+10\cos(2\varphi^{\ast}-2\ell-2\omega)
+9\cos(2\varphi^{\ast}+2\ell)
+9\cos(2\varphi^{\ast}-2\ell)
-34\cos(2\varphi^{\ast}-4\ell-2\omega)\big)\nonumber \\
&&
+\frac{1}{2}QSe^2\big(
6\sin \varphi^{\ast}
+9\sin(\varphi^{\ast}+2\ell)
+9\sin(\varphi^{\ast}-2\ell)
+10\sin(\varphi^{\ast}-2\ell-2\omega)
-34\sin(\varphi^{\ast}-4\ell-2\omega)\nonumber
\big)\\
&&
+\frac{1}{2}QS^3\big(
-2\sin \varphi^{\ast}
+\sin(\varphi^{\ast}+2\ell+2\omega)\nonumber
+\sin(\varphi^{\ast}-2\ell-2\omega)
\big)\\
&&
+\frac{1}{8}S^4\big(
\cos(2\varphi^{\ast}+2\ell+2\omega)\nonumber
-\cos(2\varphi^{\ast}-2\ell-2\omega)\big)
+\frac{1}{2}Q^2 S^2\big(1-\cos(2\ell+2\omega)\big)\Big]
\end{eqnarray}
where we have adopted the approximation $P^2=\sin^2\theta^{\ast} \sim 1$ when this quantity appears multiplied by a third or fourth-degree term, because $(1-P^2)=\cos^2\theta^{\ast}$ is of the order of $S^2$.

The terms of third and fourth degree of $U_{\rm lag}$ are:

\begin{eqnarray}
\delta_3 U_{\rm lag}&=&-\frac{3k_dGMR^{5}}{8a^3r^{\ast 3}}
\Big{[}
\frac{1}{12}e^3(27\epp_5\sin\ell+53\epp_{18}\sin 3\ell)
+\frac{1}{8}e^3\big(
\epp_2\sin(2\varphi^{\ast}-\ell-2\omega)\\
&& \phantom{mmmmmmm}
-123\epp_1\sin(2\varphi^{\ast}-3\ell-2\omega)
+\frac{845}{3}\epp_{19}\sin(2\varphi^{\ast}-5\ell-2\omega)
+\frac{1}{3}\epp_{20}\sin(2\varphi^{\ast}+\ell-2\omega)\big)\nonumber \\
&&
+\frac{1}{2}eS^2\big(
-6\epp_5\sin \ell   
-\epp_{12}\sin(\ell+2\omega)
+7\epp_{13}\sin(3\ell+2\omega)
+3\epp_{10}\sin(2\varphi^{\ast}+\ell)
+3\epp_{11}\sin(2\varphi^{\ast}-\ell)\nonumber \\
&&\phantom{mmmm}
+\epp_2\sin(2\varphi^{\ast}-\ell-2\omega)
-7\epp_1\sin(2\varphi^{\ast}-3\ell-2\omega)\big)\nonumber \\
&&
-SQe \big(
\epp_{16}\cos(\varphi^{\ast}-\ell-2\omega)
-7\epp_{17}\cos(\varphi^{\ast}-3\ell-2\omega)
+3\epp_{14}\cos(\varphi^{\ast}+\ell)
+3\epp_{15}\cos(\varphi^{\ast}-\ell)\big)\Big{]}\nonumber
\end{eqnarray}
and
\begin{eqnarray}
\delta_4 U_{\rm lag}&=&-\frac{3k_dGMR^{5}}{8a^3r^{\ast 3}}\Big{[}
\frac{7}{3}e^4 (\epp_6\sin 2\ell+\frac{11}{4}\epp_{22}\sin 4\ell)
+\frac{1}{12}e^4\big(\frac{39}{2}\epp_0\sin(2\varphi^{\ast}-2\ell-2\omega)
\\
&&\phantom{mmmmmmm}
+\epp_{31}\sin(2\varphi^{\ast}+2\ell-2\omega)
-460\epp_{3}\sin(2\varphi^{\ast}-4\ell-2\omega)
+\frac{1599}{2}\epp_{29}\sin(2\varphi^{\ast}-6\ell-2\omega) \big)\nonumber \\
&&
+\frac{1}{4}S^2e^2\big(
-18\epp_6\sin 2\ell
-10\epp_7\sin(2\ell+2\omega)
+6\epp_4\sin 2\varphi^{\ast}
+10\epp_0\sin(2\varphi^{\ast}-2\ell-2\omega)\nonumber \\
&&\phantom{mmmmm}
+34\epp_{21}\sin(4\ell+2\omega)
+9\epp_{27}\sin(2\varphi^{\ast}+2\ell)
+9\epp_{28}\sin(2\varphi^{\ast}-2\ell)
-34\epp_3\sin(2\varphi^{\ast}-4\ell-2\omega)\big)\nonumber \\
&&
+\frac{1}{2}QSe^2\big(
-6\epp_9\cos \varphi^{\ast}
-9\epp_{24}\cos(\varphi^{\ast}+2\ell)
-9\epp_{25}\cos(\varphi^{\ast}-2\ell)
+34\epp_{23}\cos(\varphi^{\ast}-4\ell-2\omega)
\nonumber \\
&&\phantom{mm}
-10\epp_8\cos(\varphi^{\ast}-2\ell-2\omega)\big)
+\frac{1}{2}QS^3\big(
2\epp_9\cos \varphi^{\ast}
-\epp_{26}\cos(\varphi^{\ast}+2\ell+2\omega)
-\epp_8\cos(\varphi^{\ast}-2\ell-2\omega)\big)\nonumber \\
&&
+\frac{1}{8}S^4  \big( \epp_{30}\sin(2\varphi^{\ast}+2\ell+2\omega)
-\epp_0\sin(2\varphi^{\ast}-2\ell-2\omega)\big)
-\frac{1}{2}\epp_7Q^2 S^2\sin(2\ell+2\omega)\Big{]}\nonumber 
\end{eqnarray}

In these expansion several new lags $\ep_j$ appear. They are not individually identified (except for those given in Appendix B). By definiton, they are the lags of the arguments of the terms in which they appear (see Sect. 4). For instance, in the term $[...]\epp_{11}\sin(2\varphi^{\ast}-\ell)$,\ $\ep_{11}$ is the lag of the argument $2\varphi^{\ast}-\ell$. 

The third-degree components of the forces are
 
\begin{eqnarray}
\delta_3 F_{1}&=&
-\frac{9k_d G M^2 R^{5}}{8a^7}\Big{[}
e^3\big(
-31\epp_0
+\frac{781}{8}\epp_1
+\frac{11}{8}\epp_2
-68\epp_3
+\frac{5}{4}\epp_5
+6\epp_6\big)\sin\ell 
\\&&\phantom{mmm}
+eS^2\big(
-8\epp_0
+7\epp_1
+\epp_2
-6\epp_5
+8\epp_8
+3\epp_{14}
-3\epp_{15}
-\epp_{16}
-7\epp_{17}\big)\sin\ell   \nonumber
\\   & &\phantom{mmm}
+e^3\big(
\frac{143}{3}\epp_0
-\frac{7}{4}\epp_1
-\frac{43}{4}\epp_2
+7\epp_5
+6\epp_6
+\frac{53}{12}\epp_{18}
-\frac{845}{24}\epp_{19}
+\frac{1}{24}\epp_{20}\big)\sin 3\ell \nonumber
\\&&\phantom{mmm}
+eS^2\big(
-2\epp_0
+\frac{1}{2}\epp_2
-\frac{3}{2}\epp_5
+2\epp_7
+4\epp_8
+\frac{3}{2}\epp_{11}
-\frac{1}{2}\epp_{12}
-3\epp_{15}
-\epp_{16}\big)\sin(\ell+2\omega)\nonumber
\\&&
+eS^2\big(
-2\epp_0
-\frac{7}{2}\epp_1
+4\epp_4
+\frac{3}{2}\epp_5
+2\epp_7
+4\epp_8
-8\epp_9
+\frac{3}{2}\epp_{10}
+\frac{7}{2}\epp_{13}
-3\epp_{14}
+7\epp_{17}\big)\sin(3\ell+2\omega)\Big] \nonumber
\end{eqnarray}
\begin{eqnarray}
\delta_3 F_{2}&=&
\frac{3k_d G M^2 R^{5}}{8a^7}\Big{[} 
2e^2S\big(
8\epp_0
-7\epp_1
-\epp_2
+6\epp_5
-4\epp_8
+4\epp_9
+3\epp_{14}
+9\epp_{15}
+\epp_{16}
-21\epp_{17}\big)\cos(\ell+\omega)
\\&&\phantom{mmm}
+e^2S\big(
13\epp_0
+7\epp_1
-3\epp_2
-17\epp_3
+6\epp_5
+9\epp_6
-\frac{53}{2}\epp_8
+\frac{11}{2}\epp_9
+6\epp_{15}
+6\epp_{16}
\big)\cos(\ell-\omega) \nonumber
\\&&\phantom{mmm}
-e^2S\big(
+29\epp_0
-7\epp_1
-5\epp_2
-17\epp_3
+18\epp_5
+9\epp_6
+\frac{11}{2}\epp_8
-\frac{53}{2}\epp_9
-18\epp_{14}
+14\epp_{17}
\big)\cos(3\ell+\omega) \nonumber
\\&&\phantom{mmm}
-S^3\big(
\epp_0
-\epp_4
-\epp_7
-\frac{5}{2}\epp_8
+\frac{3}{2}\epp_9
\big)\big(\cos(\ell+\omega)-\cos(3\ell+3\omega)\big)\Big] \nonumber
\\&+&
\frac{3k_d G M^2 R^{5}}{8a^7}\Big{[}
9e^2S\epp_{25}\cos(\ell-\omega)+2e^2S(3\epp_9+5\epp_8)\cos(\ell+\omega)
+S^3(2\epp_7+\epp_8-2\epp_9)\cos(\ell+\omega)
\nonumber
\\ && \phantom{mmm}
-e^2S(34\epp_{23}-9\epp_{24})\cos(3\ell+\omega) 
-S^3(2\epp_7-\epp_{26})\cos(3\ell+3\omega)\Big]
 \nonumber
\end{eqnarray}
\begin{eqnarray}
\delta_3 F_{3}&=&
-\frac{3k_d G M^2 R^{5}}{8a^7}\Big{[}   
e^3\big(
+58\epp_0
-\frac{423}{4}\epp_1
-\frac{7}{4}\epp_2
-136\epp_3\big)\cos\ell
\\&&\phantom{mmm}
+eS^2\big(
12\epp_0
+\frac{21}{2}\epp_1
-\frac{3}{2}\epp_2
-8\epp_8
-8\epp_9
-3\epp_{14}
-3\epp_{15}
+\epp_{16}
-7\epp_{17}
\big)\cos\ell
\\&& \nonumber\phantom{mmm}
-e^3\big(
+98\epp_0
+\frac{7}{2}\epp_1
-\frac{43}{2}\epp_2
+\frac{845}{12}\epp_{19}
+\frac{1}{12}\epp_{20}\big)\cos 3\ell \nonumber
\\&& \nonumber\phantom{mmm}
+eS^2\big(
-4\epp_0
+\frac{7}{4}\epp_1
+\frac{3}{4}\epp_2
+4\epp_8
-3\epp_{11}
+3\epp_{15}
-\epp_{16}\big)\cos(\ell+2\omega)
\\&& \nonumber\phantom{mmm}
+eS^2\big(
-\frac{21}{4}\epp_1
-\frac{1}{4}\epp_2
-8\epp_4
+4\epp_8
+8\epp_9
-3\epp_{10}
+3\epp_{14}
+7\epp_{17}\big)\cos(3\ell+2\omega) \Big]
\end{eqnarray}

In the expression of $\delta_3 F_{2}$, one may note two parts graphically separated. The first of them comes from $\delta_3 U_{\rm lag}$ and the second from $\delta_4 U_{\rm lag}$. In the same way, $\delta_3 U_{\rm lag}$ contributes terms of degree 2 to $F_2$, which were included in eqn. \ref{f2ueva}.
 
The lost of one degree in the derivation of the force comes from derivatives with respect to $Q$ ($Q$ is of the order ${\cal O}(S)$, but its derivative $2-4 P^2 \simeq -2$ is a finite quantity). 

\section{Coriolis accelerations}

In Darwin's theory, the rotation of the tidally deformed body does not appear explicitly in the tidal forces. It only enters through the tidal lag angles and by the association of each lag to the corresponding wave in the body. For instance, when we say that $\epp_2$ corresponds to the semi-diurnal wave (in Type I tides), we are implicitly assuming the equator of the deformed body as reference plane. Since this plane is not fixed, we have to take into account the rising of apparent forces due to its motion (see e.g. Ferraz-Mello, 2007, p. 17):
\beq
\mathbf{F}_{\rm app} = M \mathbf{O} \times(\mathbf{r} \times  \mathbf{O}) + 2M \mathbf{v}\times \mathbf{O} 
\endeq
where $\mathbf{O}$ is the rotation angular velocity vector of the reference frame
The first term is the {\it centrifugal force}. It is proportional to the square of the angular rotation velocity and can be neglected at the order considered in this report. Indeed, in this case, the frame rotates around the nodal line with $|\mathbf{O}|=\dot{J}\sim {\cal O}(S\ep)$. Therefore the centrifugal force is of order ${\cal O}(S^2\ep^2)$. 
The second term is the {\it Coriolis force}. This term is of the same order of the terms considered in previous sections and, for sake of completeness, needs to be calculated.

The vectors $\mathbf{v}$ and $\mathbf{O}$ belong to the orbital plane ($\mathbf{O}$ lies on the nodal line). Therefore the Coriolis force is perpendicular to the orbital plane. 
The first conclusion is that no work is done (the scalar product by $\mathbf{v}$ is zero) and, thus, it does not affect the orbital semi-major axis. 
However, it has a torque directed along the transverse component of the orbital velocity. First we have to express $\mathbf{v}$ in a reference system whose in-plane unit vectors are $\hat{\mathbf{n}}$ along the nodal line and $\hat\mathbf{m}$ normal to it. 
The frame angular velocity vector is then $\mathbf{O} = \dot{J}\,\hat{\mathbf n}$ and the orbital velocity of $\tens{M}$ is
\beq
{\mathbf v} = \frac{na}{\sqrt{1-e^2}}\big[ 
\big(\sin(\omega+v)+e\sin{\omega}\big) \,\hat{\mathbf n} -
\big(\cos(\omega+v)+e\cos{\omega}\big) \,\hat{\mathbf m} \big].
\endeq 
Hence
\beq
\mathbf{f}_{\,\rm Cor} = 2M \mathbf{v}\times \mathbf{O} = 2M\dot{J}
\frac{na}{\sqrt{1-e^2}}\big(\cos(\omega+v)+e\cos{\omega}\big)\, \hat{\mathbf u}
\endeq
where, as before, $\hat{\mathbf u}$ is a unit vector normal to the orbital plane.
The torque due to the Coriolis force is
\beq
{\cal M}_{\,\rm Cor} = \mathbf{r}\times \mathbf{f}_{\,\rm Cor} =
2M\dot{J}
\frac{nar}{\sqrt{1-e^2}}\big(\cos(\omega+v)+e\cos\omega\big) \,\hat{\mathbf t}
\endeq
where $\hat{\mathbf t}$ is the transverse\footnote{that is, normal to the radius vector} unit vector: 
\begdi
\hat{\mathbf t}=e^{i(\omega+v)}\,\hat{\mathbf m} = -\sin(\omega+v)\,\hat{\mathbf n} +\cos(\omega+v)\,\hat{\mathbf m}.
\enddi

If the two components of the torque are averaged over one period, we obtain, respectively:
\begin{eqnarray}
M_n & \speq & \frac{1}{2}M\dot{J}na^2e^2\sin 2\omega\\ \nonumber
M_m & \speq & M\dot{J}na^2(1+\frac{3}{2}e^2) +\frac{1}{2}M\dot{J}na^2e^2\cos 2\omega
\end{eqnarray}

To remain in the order of magnitude of the previous calculations, we neglect terms in $\dot{J}e^2$. Thus ${\cal M}_{\ \rm Cor}= M\dot{J}na^2\,\hat{\mathbf m}$. If this torque is substituted in the equation $\frac{d}{dt}{\cal L} = {\cal M}$, there results a homogeneous equation in the variables $\frac{d}{dt}|{\cal L}|,\ \dot{N},\ (M\dot{J}na^2 + \dot{I}|{\cal L}|)$, where $N=\widehat{\rm {ON}}$, with the trivial solution $0,0,0$. Then $|{\cal L}|$ remains constant, that is, the eccentricity is not affected by the Coriolis force (at the considered order), $\dot{N}=0$ (no precession of the node) and 
\beq
\dot{I}_{\,\rm Cor}=-\frac{M\dot{J}na^2}{|{\cal L}|} \sim -\dot{J}
\endeq
This is the same result which would be obtained from an elementary geometric reasoning. The decrease of the obliquity $I$ is equal to the amount of the rotation of the reference plane. In other words, the orbital plane is not affected.  

The effect of the Coriolis force on the rotating mass $\tens{m}$ is similar. Elementary calculations allow us to obtain for the apparent torque, ${\cal M}_{\ \rm Cor}= -\dot{J} C \mathbf{\Omega} \times \hat{\mathbf n}$ which corresponds to counter-rotate the body around the nodal line with the same rotation angular velocity as the frame.


\begin{thebibliography}{}

\bibitem{Ale}
Alexander, M.E.: 1973, ``The weak friction approximation and tidal evolution in close binary systems", Astrophys. Sp. Sci. \textbf{23}, 459-510.

\bibitem{Beu}
Beutler, G.: 2005, Methods of Celestial Mechanics, Vol. I (Springer, Berlin)

\bibitem{Cha}
Chandrasekhar, S. 1969, Ellipsoidal Figures of Equilibrium, Chap. VIII (Yale Univ. Press, New Haven)

\bibitem{D79} 
Darwin, G.H.: 1879, ``On the precession of a viscous spheroid and on the remote history of the Earth'', Philos. Trans. \textbf{170}, 447-530 (repr. Scientific Papers, Cambridge, Vol. II, 1908).

\bibitem{D80} 
Darwin, G.H.: 1880, ``On the secular change in the elements of the orbit of a satellite revolving about a tidally distorted planet'', Philos. Trans. \textbf{171}, 713-891 (repr. Scientific Papers, Cambridge, Vol. II, 1908).

\bibitem{Dob}
Dobbs-Dixon, I., Lin,D.N.C. and Mardling, R.A.  2004
''Spin-orbit evolution of short-period Planets",  Astrophys. J. \textbf{610}, 464-476

\bibitem{Efr} 
Efroimsky, M. and Lainey, V.: 2007, ``Physics of Bodily Tides in Terrestrial Planets and the Appropriate Scales of Dynamical Evolution",
J. Geophys. Res. \textbf{112}, E12003

\bibitem{Ef8}
Efroimsky, M.: 2008
''Tidal torques. I. A critical review of some techniques", Cel. Mech. Dynam. Astron. (submitted)		
 
\bibitem{Egg}
Eggleton, P.P., Kiseleva, L.G. and Hut, P.: 1998, ``The Equilibrium Tide Model for Tidal Friction", Astrophys. J. \textbf{499}, 853-870.

\bibitem{SFM}
Ferraz-Mello, S.:2007, ``Canonical Perturbation Theories. Degenerate Systems and Resonance" (Springer, New York)

\bibitem{Gol} 
Goldreich, P. 1963. ``On the eccentricity of satellite orbits in the Solar System". {Mon. Not. R. Astron. Soc} \textbf{126}, 257-268.

\bibitem{G66}
Goldreich, P. 1966. ``Final spin states of planets and satellites". Astron. J. \textbf{71}, 1-7.

\bibitem{GSo}
Goldreich, P. and Soter, S.: 1966, ``Q in the Solar System", Icarus, \textbf{5} 375-389.
 
\bibitem{Hut}
Hut, P.: 1981, ``Tidal evolution in close binary systems", Astron. Astrophys. \textbf{99}, 126-140.

\bibitem{Jea}
Jeans, J. 1928, Astronomy and Cosmogony, Sec. 215-216 (CUP, Cambridge). (repr: Dover, New York, 1961)

\bibitem{Jef} 
Jeffreys, H. 1961. ``The effect of tidal friction on eccentricity and inclination". {Mon. Not. R. Astron. Soc} \textbf{122}, 339-343.

\bibitem{Kau}
Kaula, W.M. 1964. ``Tidal dissipation by solid friction and the resulting orbital evolution",
Rev. Geophys. \textbf{3} 661-685.
 
\bibitem{Lem}
Lemaitre, A., D'Hoedt, S. and Rambaux, N.: 2006,	
``The 3:2 Spin-Orbit Resonant Motion of Mercury'' Cel. Mech. Dynam. Astron.	
\textbf{95}, 213-224	

\bibitem{Lev}
Levrard, B.: 2008, ``A proof that tidal heating in a synchronous rotation is always larger than in an asymptotic nonsynchronous rotation state", Icarus \textbf{193}, 641-643

\bibitem{Lov}
Love, A.E.H.: 1927. ``A Treatise on the Mathematical Theory of Elasticity", CUP, Cambridge. 

\bibitem{Mac} 
MacDonald, G.F.: 1964, ``Tidal Friction", Rev. Geophys. \textbf{2}, 467-541.

\bibitem{Mar}
Mardling, R.A. and Lin, D.N.C.: 2004
''On the Survival of Short-Period Terrestrial Planets",  Astrophys. J. \textbf{614}, 955-959

\bibitem{Mi9} 
Mignard, F.: 1979, The evolution of the lunar orbit revisited - I, Moon and Planets 
\textbf{20}, 301-315.

\bibitem{M80} 
Mignard, F.: 1980, The evolution of the lunar orbit revisited - II, Moon and Planets \textbf{23}, 185-201.  

\bibitem{Mun}
Munk,W.H. and MacDonald, G.J.F.: 1960, The rotation of the Earth (CUP, Cambridge).

\bibitem{Pea}
Peale, S.J., Cassen, P. and Reynolds, R.T.: 1980, ``Tidal dissipation, orbital evolution and the nature of Saturn's inner satellites", Icarus, \textbf{43}, 65-72.

\bibitem{Sea} Sears, W.D., Lunine, J.I. and Greenberg, R.:1993, "Equilibrium nonsynchronous rotation of Titan", Icarus, \textbf{105}, 259-262.

\bibitem{Seg}
Segatz, M., Spohn, T., Ross, M.N. and Schubert, G.: 1988, ``Tidal dissipation, surface heat flow and figure of viscoelastic models of Io", Icarus, \textbf{75}, 187-206.

\bibitem{Tis}
Tisserand, F.: 1891, Trait\'e de M\'ecanique C\'eleste, tome II, chap. VIII
(Gauthier-Villars, Paris).

\bibitem{Wil}
Williams, J.G., Turyshev, S.G., Boggs, D.H. and Ratcliff, J.T.: 2006, ``Lunar laser ranging science: Gravitational Physics and Lunar Interior and Geodesy", Advances in Space Research, 37, 67-71.

\bibitem{Win}
Winn, J.N. and Holman, M.J.: 2005, ``Obliquity tides on hot Jupiters" \textbf{628}, L159-L162.

\bibitem{Wis}
Wisdom, J.: 2004, ``Spin-orbit secondary resonance dynamics of Enceladus", Astron. J. \textbf{128}, 484-491.

\bibitem{Wi7}
Wisdom, J.: 2008, ''Tidal dissipation at arbitrary eccentricity and inclination", Icarus \textbf{193}, 637-640.

\bibitem{Yod}
Yoder, C.F. and Peale, S.J.: 1981, ``The tides of Io", Icarus \textbf{47}, 1-35.

\bibitem{Zah}
Zahn, J.P.: 1977, ``Tidal friction in close binary stars", Astron. Astrophys. 
\textbf{57}, 383-394.

\end{thebibliography}
\end{document}